\documentclass[twocolumn,nobibnotes,showpacs,preprintnumbers,amsmath,amssymb]{revtex4}
\pdfoutput=1
\usepackage{graphicx}% Include figure files
\usepackage{dcolumn}% Align table columns on decimal point
\usepackage{bm}% bold math
\newcommand{\be}{\begin{eqnarray}}
\newcommand{\ee}{\end{eqnarray}}
\newcommand{\bea}{\begin{eqnarray}}
\newcommand{\eea}{\end{eqnarray}}

\newcommand{\keV}{{~\rm keV}}
\newcommand{\MeV}{{~\rm MeV}}
\newcommand{\GeV}{{~\rm GeV}}

\newcommand{\mX}{m_{\scriptstyle \chi} }
\newcommand{\mN}{m_{\scriptstyle N}}

\newcommand{\ER}{E_{\scriptstyle R}}

\newcommand{\sDD}{\sigma_{\rm \scriptstyle DD}}
\newcommand{\sDZ}{\sigma_{\rm \scriptstyle DZ}}

\newcommand{\darkmed}{A^\prime}

\begin{document}

\title{Magnetic Inelastic Dark Matter}
\author{Spencer Chang$^{(a)}$, Neal Weiner$^{(b)}$, and Itay Yavin$^{(b)}$}

\affiliation{(a) Physics Department,~University of California Davis, Davis,~California 95616 \\ (b) Center for Cosmology and Particle Physics, Department of Physics, New York University, New York, NY 10003}

%\date{\today}

\begin{abstract}
Iodine is distinguished from other elements used in dark matter direct detection experiments both by its large mass as well as its large magnetic moment. Inelastic dark matter utilizes the large mass of iodine to allay tensions between the DAMA annual modulation signature and the null results from other experiments. We explore models of inelastic dark matter that also take advantage of the second distinct property of iodine, namely its large magnetic moment. In such models the couplings are augmented by magnetic, rather than merely electric, interactions. These models provide simple examples where the DAMA signal is compatible with all existing limits. We consider dipole moments for the WIMP, through conventional magnetism as well as ``dark'' magnetism, including both magnetic-magnetic and magnetic-electric scattering. We find XENON100 and CRESST should generically see a signal, although suppressed compared with electric inelastic dark matter models, while KIMS should see a modulated signal comparable to or larger than that of DAMA. In a large portion of parameter space, de-excitation occurs promptly, producing a $\sim 100$ keV photon inside large xenon experiments alongside the nuclear recoil. This effect could be searched for, but if not properly considered may cause nuclear recoil events to fail standard cuts.
\end{abstract}

\maketitle

\section{Introduction}
The DAMA annual modulation signature has persisted and grown to 8.9 $\sigma$~\cite{Bernabei:2010mq}. The constraints imposed by null results from many sources \cite{Aprile:2010um,Ahmed:2009zw,Sanglard:2009qp,Lang:2009ge,Lebedenko:2009xe} would seem to exclude a dark matter interpretation of the signal within the framework of conventional weakly interacting massive particles (WIMPs) scattering with spin-independent interactions. In this context, a wide range of proposals have been put forward to explain the apparent tension, such as spin-coupled WIMPs \cite{Ullio:2000bv,Belli:2002yt,Savage:2004fn,Bernabei:2005hj,Fairbairn:2008gz}, mirror matter \cite{Foot:2004gh,Foot:2008nw}, resonant dark matter \cite{Bai:2009cd}, exothermic dark matter \cite{Essig:2010ye,Graham:2010ca}, momentum-dependent scattering \cite{Feldstein:2009tr,Chang:2009yt},  and light WIMPs with or without ion channeling \cite{Belli:1999nz,Bottino:2003cz,Bernabei:2007hw,Bottino:2008mf,Chang:2008xa,Savage:2008er,Petriello:2008jj,Hooper:2010uy}.   

Inelastic dark matter (iDM) \cite{TuckerSmith:2001hy} was originally proposed to explain DAMA not through a change in couplings, but rather through a change in kinematics. It brought into focus the large mass difference between iodine (DAMA) and germanium (CDMS). The iDM framework has three basic elements
\begin{itemize}
\item{The existence of an excited WIMP state $\chi^*$, with $\delta = m_{\chi^*}-m_\chi \sim \mu_{\scriptstyle{N\chi}} v^2$, where $\mu_{\scriptstyle{N\chi}}$ is the reduced WIMP-nucleus mass and $v$ is the WIMP velocity.}
\item{An allowed inelastic scattering against the nucleus $\chi + N \longrightarrow \chi^* + N$.}
\item{A forbidden, or highly suppressed elastic scattering $\chi + N \longrightarrow \chi + N$.}
\end{itemize}
Inelastic scattering alters the minimum velocity required for a dark matter particle to scatter a nuclei $N$ with recoil energy $E_R$,
\begin{equation}
v_{\rm min} = \frac{1}{\sqrt{2\mN\ER}}\left(\frac{\mN\ER}{\mu_{\scriptstyle{N\chi}}} + \delta \right).
\end{equation}
These kinematics change the spectrum of the signal (shifting it to higher recoil energy), favor heavy targets (such as DAMA's iodine) over lighter targets (such as CDMS's germanium), and enhance the modulated signal compared to the unmodulated part (favoring modulation experiments). Together, these features allow a signal at DAMA, while highly suppressing or eliminating the signals at other experiments \cite{WeinerKribs,MarchRussell:2008dy}.

In most models of iDM \cite{TuckerSmith:2001hy, ArkaniHamed:2008qn, Baumgart:2009tn, Cui:2009xq, Cheung:2009qd, Katz:2009qq, Finkbeiner:2009mi, Alves:2009nf, Morrissey:2009ur, Anber:2009tz, Arina:2009um, Chen:2009ab, Kaplan:2009de, Kumar:2009sf,Chang:2010pr} scattering arises dominantly through a vector current, such as a the Z-boson, or a new, light vector. However, many interactions can mediate inelastic transitions. For instance, recently \cite{Kopp:2009qt} argued that a coupling to the proton's spin would significantly weaken constraints from existing experiments. While standard vector interactions are still viable, the strong constraints from ZEPLIN-III~\cite{Lopes:2010zz} and CRESST~\cite{Lang:2009ge}, in particular, require high rates of modulation and departures from a Maxwellian halo \cite{Kuhlen:2009vh,Lang:2010cd, Alves:2010pt} (although such departures are expected when sampling the tail of the distribution \cite{Vogelsberger:2008qb,Kuhlen:2009vh}).

In this short letter, we point out that DAMA's iodine target also distinguishes itself from other detector's targets through its magnetic properties. Iodine's large nuclear magnetic moment compels us to consider models with dominantly {\em magnetic} interactions as a way to reconcile DAMA's positive detection claims with the other null results. We concentrate on dipole-dipole scattering, which can arise from conventional magnetism or a dipole of a dark force. In both cases, we shall see parameter space opens, with inelastic dipole-dipole scattering essentially unconstrained.

\section{Magnetic Inelastic Dark Matter}
If one wants to understand how DAMA could have a positive signal while other experiments do not, there are many directions one can pursue. Narrowing the focus on nuclear recoils induced by WIMP collisions, we must examine what the differences are between NaI and the other existing targets.

The original iDM proposal focused on a single dimension, namely the kinematical properties of iodine. As it is much heavier than many targets, in particular germanium, this allowed a significant departure from conventional elastic expectations. The fact that DAMA focuses on relatively high energies ($\sim 20+$ $\rm keV_R$ off iodine assuming the standard quenching factor $q_I = 0.08$) and modulation gave additional changes when comparing to elastic scattering limits, but ultimately the key distinction was the kinematical change of a heavy target.

This simple one-dimensional analysis is important, but iodine's {\em magnetic} properties also distinguish it from most other target nuclei.
\begin{figure}
\begin{center}
\includegraphics[width=0.45 \textwidth]{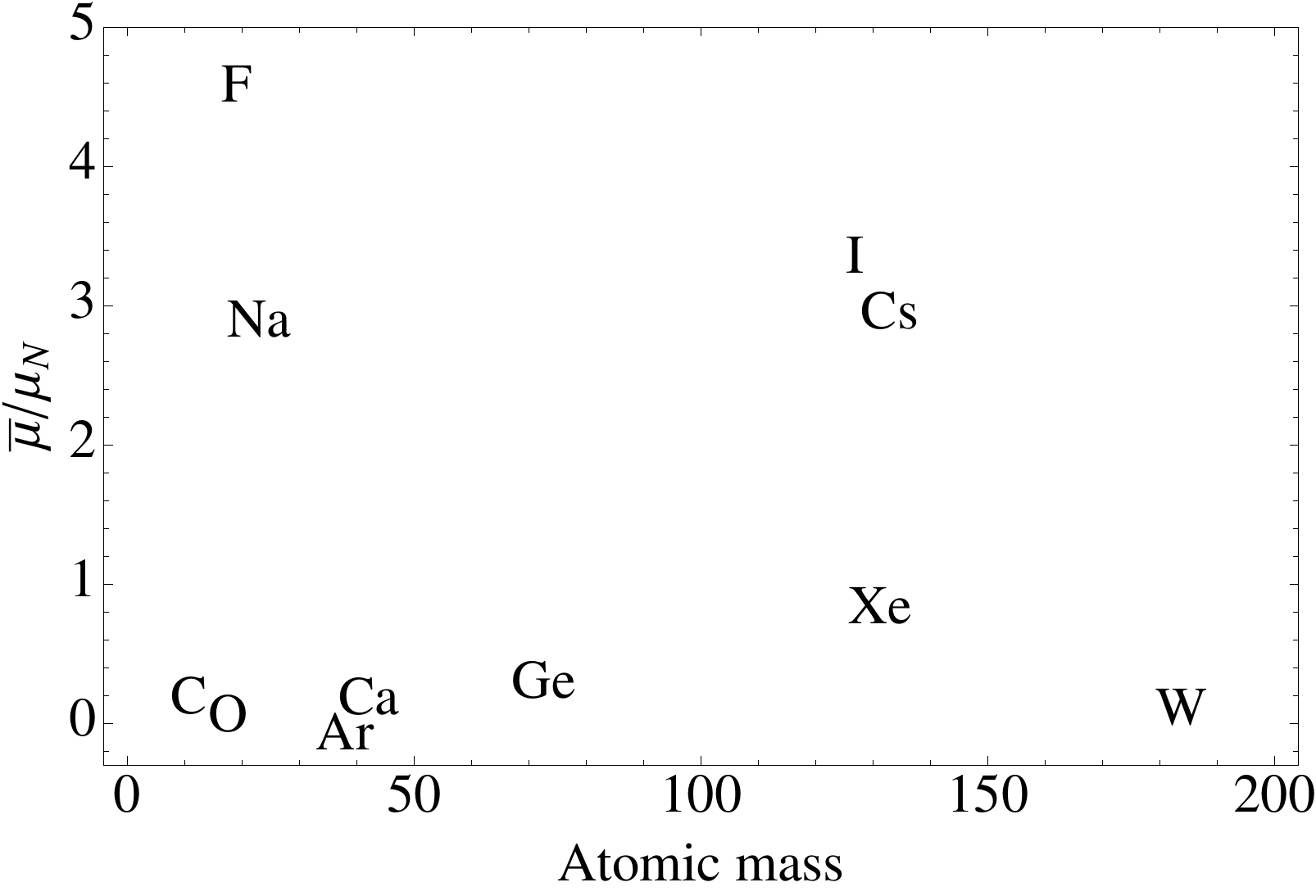}
\end{center}
\caption{The weighted-atomic mass and weighted-magnetic dipole moment (Eq.~(\ref{eqn:weighteddipole}) in units of the nuclear magneton $\mu_{\scriptstyle{N}}$ of various dark matter search targets. (C,O and Ca,Ar have been shifted slightly so as not to overlay each other.)}
\label{fig:massanddipole}
\end{figure}
The quantity that we will see is most relevant is the weighted dipole moment
\be
\label{eqn:weighteddipole}
\bar \mu = \left( \sum_{\rm isotope} f_i \mu_i^2 \frac{S_i+1}{S_i} \right)^{1/2},
\ee
where $f_i$, $\mu_i$, and $S_i$ are the elemental abundance, nuclear magnetic moment, and spin, respectively,  of isotope $i$. We show in Fig.~\ref{fig:massanddipole} the abundance-weighted atomic masses, and the weighted dipole moment of various target nuclei. We see that while tungsten (W) has a large mass, its magnetic moment is rather small. Fluorine (F) and sodium (Na) have large magnetic dipoles but are very light. Xenon (Xe) has a couple of isotopes with appreciable dipoles, however, they are insufficient to make it competitive with iodine. The combination of large mass and large dipole makes the iodine target used by DAMA quite unique among the nuclear targets, with only KIMS'~\cite{Kim:2008zzn} cesium (Cs) target similar in its qualitative features.  The iodine dipole arises dominantly from the angular momentum of unpaired protons \cite{Ressell:1997kx}, with additional contributions from the neutron and proton spin. 

We are therefore led to consider models that make both  kinematical {\em and magnetic} distinctions between targets. Since its proposal, the focus of iDM model building has dominantly been on {\em electrically} coupled WIMPs, either directly to charge, or to some combination of the mass number $A$ and the atomic number $Z$, such as through the $Z^0$-boson. Since we wish to take advantage of the large magnetic dipole of iodine, we instead focus on models of magnetically-coupled inelastic dark matter (MiDM).

\section{Scenarios for MiDM}
The magnetic interactions of a WIMP can appear at different orders in the multipole expansion. The first order, namely a magnetic monopole, is interesting but problematic \footnote{The flux of Standard Model monopoles is strongly bounded~\cite{Amsler:2008zzb}. A monopole under some other, broken $U(1)$ has to be confined, which complicates the analysis, and we leave it for future work.}. Instead we choose to focus on the case of a magnetic dipole which has a sizable interaction with the magnetic dipole of iodine. However, a magnetically interacting WIMP also feels a velocity-suppressed interaction with the charge of the nucleus, thus one cannot simply consider scattering off magnetic moments. For iodine the contribution from $Z^2 v^2$ is subdominant to $\mu^2$, but for magnetically-challenged nuclei, such as W, or even Xe, the charge coupling can dominate the scattering.

\subsection{Dipole-Dipole Inelastic Scattering}

The idea that the WIMP could have a magnetic dipole has been long studied (see., e.g., \cite{Bagnasco:1993st,Pospelov:2000bq,Sigurdson:2004zp,Gardner:2008yn, Cho:2010br,An:2010kc}.) The dipole operator is naturally off-diagonal~\cite{Dreiner:2008tw,Kopp:2009qt}, and mediates transitions between the ground state $\chi$ and the excited state $\chi^*$ ,
\be
\mathcal{L}\supset \left(\frac{\mu_\chi}{2}\right)\bar \chi^* \sigma_{\mu\nu} F^{\mu\nu} \chi + c.c.
\ee 
where $\mu_\chi$ is the dipole strength and $\sigma_{\mu\nu} = i[\gamma_\mu,\gamma_\nu]/2$. \cite{Sigurdson:2004zp} considered such transitions in the early universe for dark matter in the range of ${\rm few\, keV}-{\rm few\, MeV}$.
\cite{Masso:2009mu} considered inelastic WIMP dipole-nuclear charge scattering to explain DAMA. Such an interaction, however, does not significantly change the relative strength of the various experiments compared with charge-charge (i.e., vector current) interactions, and the viability of the scenario found in \cite{Masso:2009mu} was largely because the significant constraints from the CRESST experiment were ignored. \cite{Kopp:2009qt} considered a related idea, studying the parameter space under the assumption of an iDM that couples to proton nuclear spin exclusively, although no particle physics model generating the required interaction was found.

Here we study inelastic dipole-dipole (DD) interactions, which occur in addition to dipole-charge interactions, and we shall see that this opens up significant parameter space. In fact, for WIMP-iodine scattering DD cross section dominates over the dipole-charge (DZ) scattering. DZ scattering, on the other hand, may dominate the scattering in other targets (such as tungsten) and provide signals at other experiments. The direct detection scattering rate is given by, 
\be
\label{eqn:QEDformula}
\frac{d \sigma}{d \ER} = \frac{d\sDD}{d \ER}+\frac{d \sDZ}{d \ER}
\ee
\begin{eqnarray}
\nonumber
\frac{d \sDD}{d \ER}&=& \frac{16 \pi \alpha^2  \mN}{v^2} \left(\frac{\mu_{nuc}}{e}\right)^2 \left(\frac{\mu_\chi}{e}\right)^2 \\[-.2cm]
\\[-.2cm]
\nonumber
 &\times& \left(\frac{S_\chi+1}{3 S_\chi}\right) \left(\frac{S_N+1}{3 S_N }\right)F_D^2[\ER]
\end{eqnarray}
 \begin{eqnarray}
\nonumber
\frac{d \sDZ}{d \ER}&=& \frac{4\pi Z^2 \alpha^2}{\ER}\left(\frac{\mu_\chi}{e}\right)^2 \bigg[1-\frac{E_R}{v^2}\left(\frac{1}{2m_N}+\frac{1}{m_\chi}\right)\\[-.15cm]
\\[-.15cm]
\nonumber
&-&\frac{\delta}{v^2}\left(\frac{1}{\mu_{\scriptstyle{N\chi}}}+\frac{\delta}{2m_N E_R}\right)\bigg]
\left(\frac{S_\chi+1}{3S_\chi}\right) F^2[\ER]
\end{eqnarray}
where $F^2[\ER]$ is the usual nuclear form-factor and $F_D^2[\ER]$ is the nuclear \textit{magnetic dipole} form-factor.  In the above expressions we have kept the leading terms in the expansion of $E_R, \delta, {\rm and}~ v$. We note the inelastic enhancement of the destructive interference in the DZ cross section which leads to a further reduction of the signal in experiments which utilize a magnetically-challenged targets.  We have checked that our formulae agree in the elastic limit with those in \cite{Bagnasco:1993st,Barger:2010gv}. 

\begin{figure*}
\includegraphics[width=5.cm]{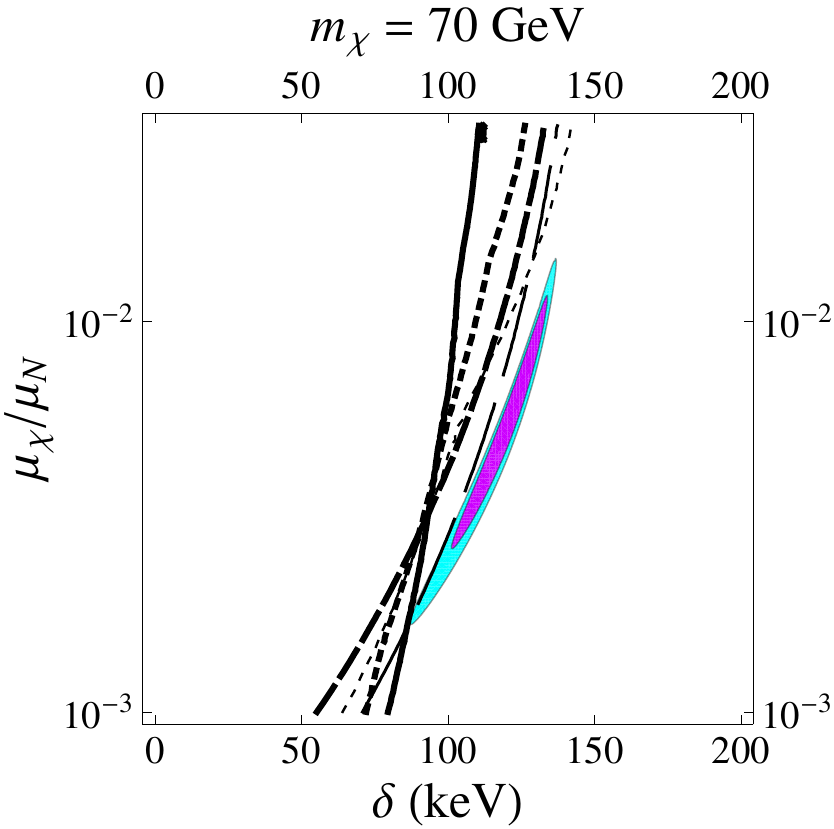}
\includegraphics[width=5.cm]{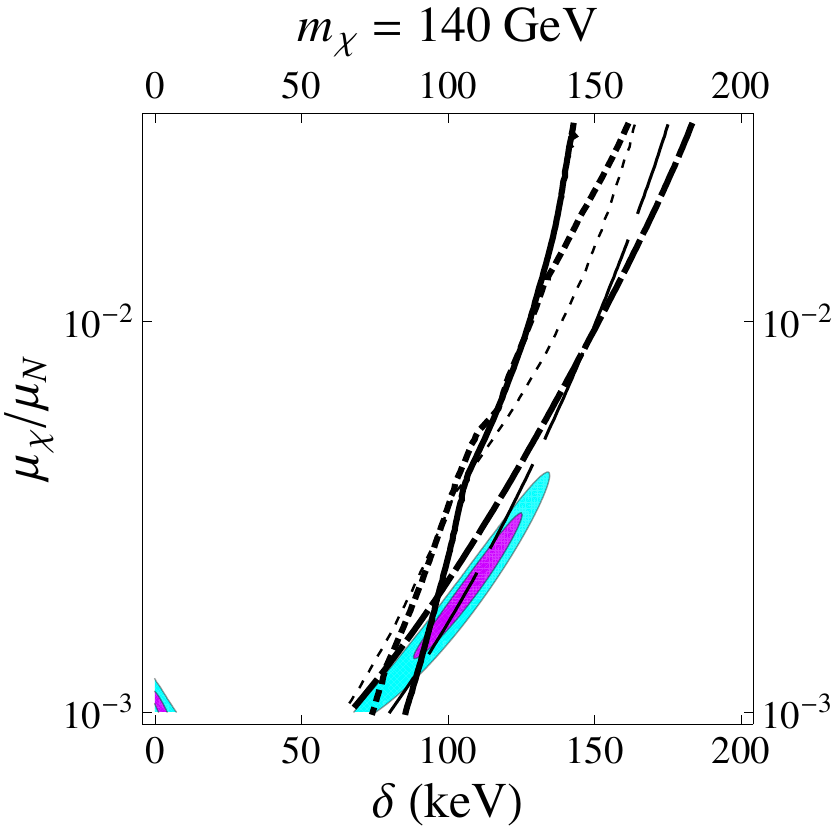}
\includegraphics[width=5.cm]{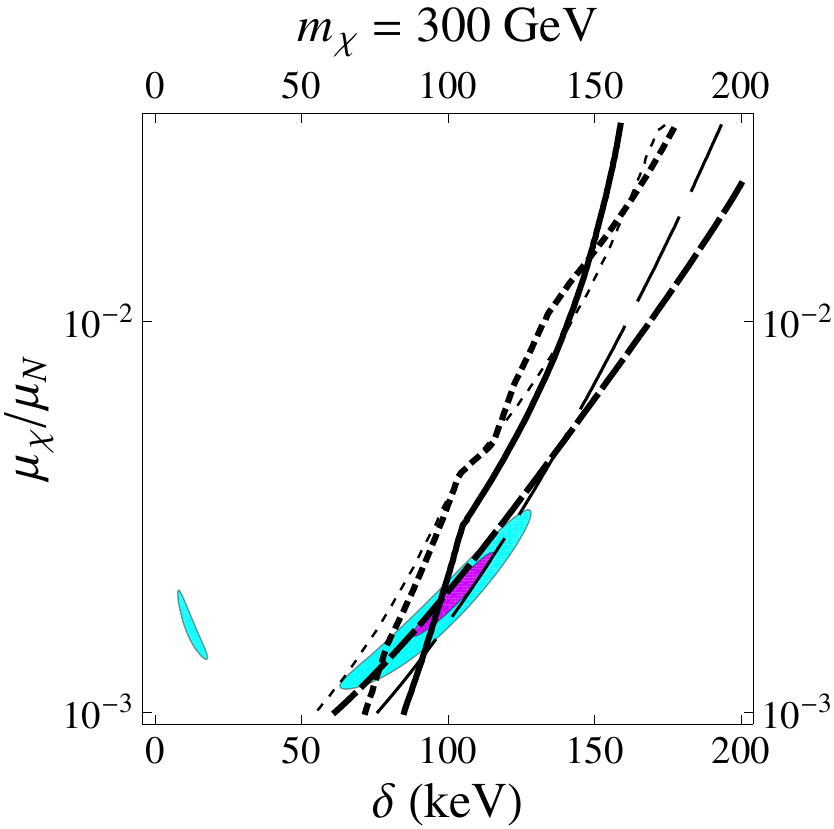}\\
\includegraphics[width=5.cm]{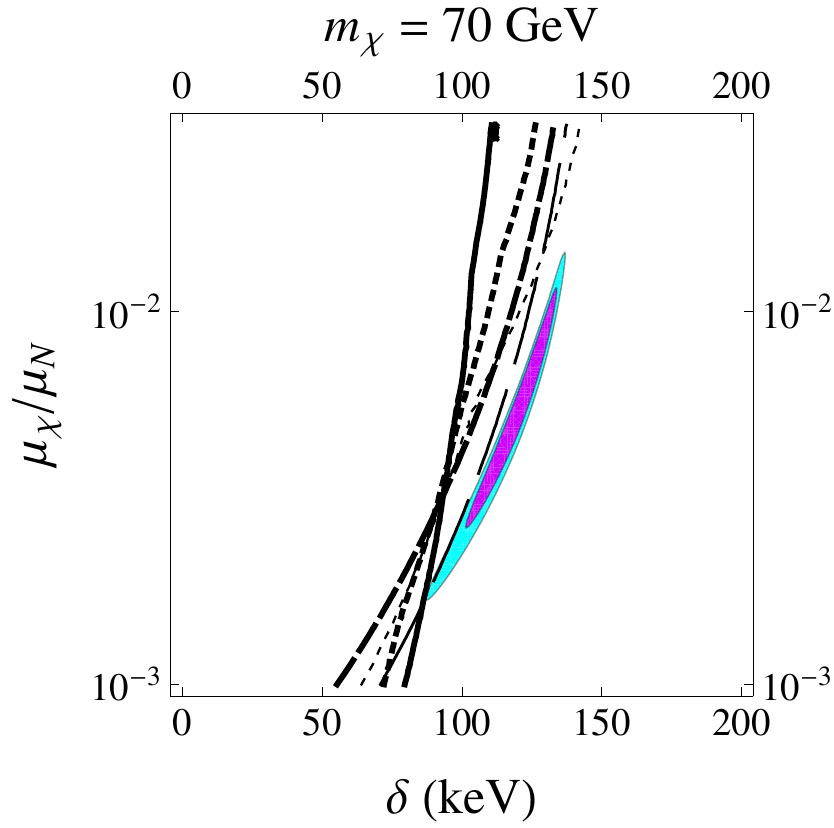}
\includegraphics[width=5.cm]{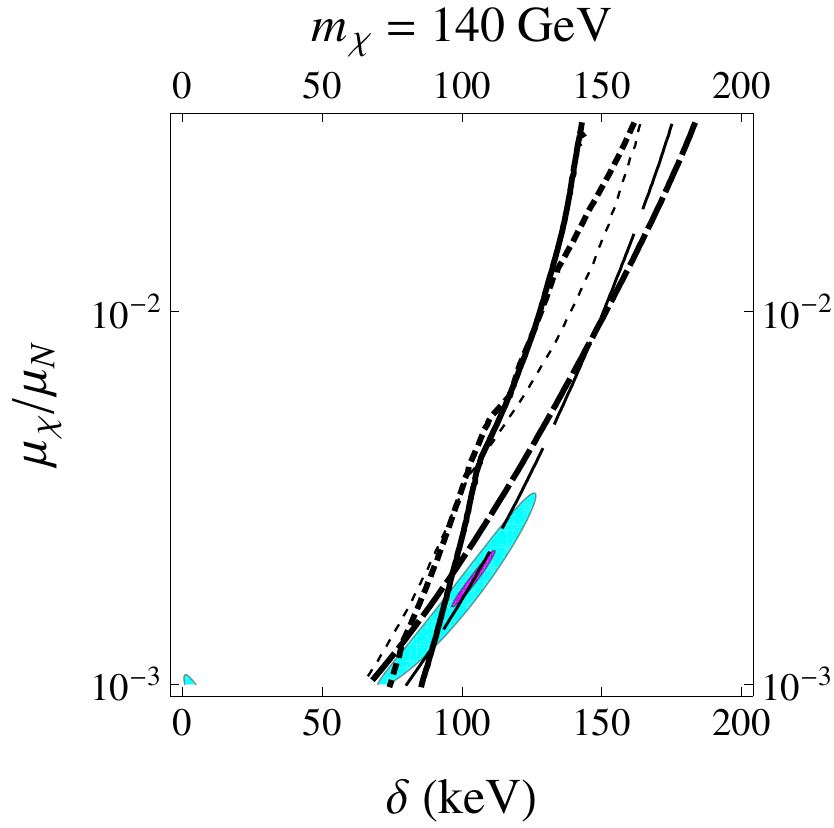}
\includegraphics[width=5.cm]{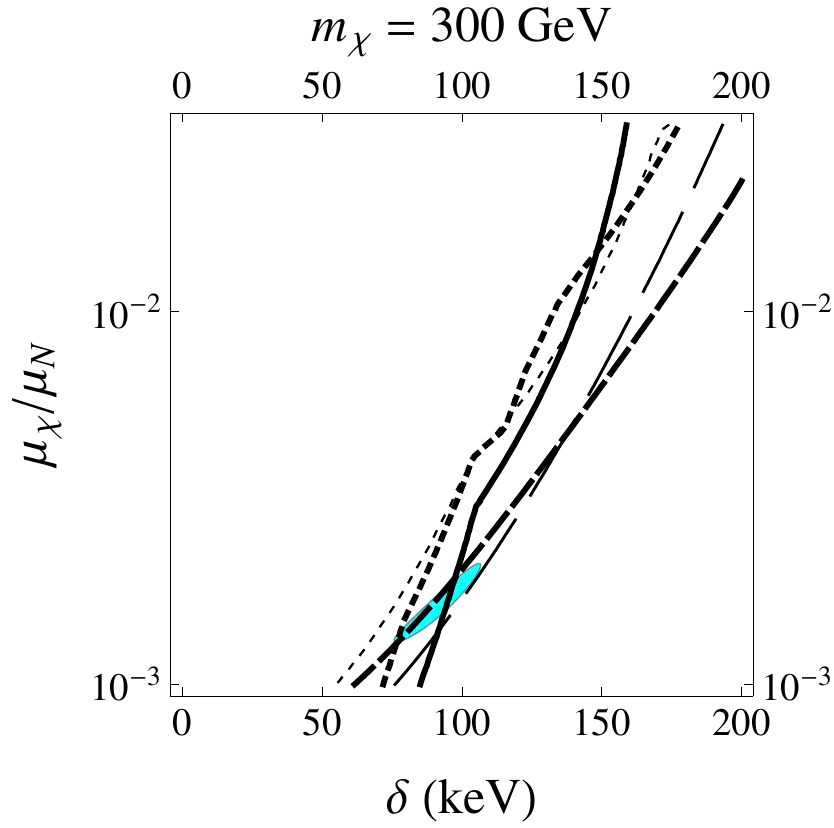}
\caption{Allowed ranges of parameter space for inelastic scattering of a magnetic dipole (dominated by dipole-dipole scattering for iodine).  Purple (dark) and blue (light) regions are 90\% and 99\% allowed confidence intervals, respectively, for a $\chi^2$ fit to the DAMA modulation. Constraints are: CDMS (solid), CRESST-II (short light dashed), XENON10 (short dark dashed), KIMS (long dark dashed), and ZEPLIN-III (long light dashed).  In the upper row, we utilized only the $2-8~$keVee bins in DAMA, and in the lower row we used the entire range of $2-14~$keVee. This serves to illustrate the strong dependence of the allowed DAMA parameter space for heavy dark matter on the dipole form factor's behavior at high $E_R$ as discussed in Sec.~\ref{sec:formfactor} in detail.
}
\label{fig:dipoledipoleregions}
\end{figure*}

We show the allowed parameter space in Fig.~\ref{fig:dipoledipoleregions}. To determine these parameter ranges, we use the binned DAMA data of \cite{Bernabei:2010mq}, taking the 2-8 keVee (upper row) and 2-14 keVee (lower row) bins and calculating a $\chi^2$ parameter for each point in $(m_\chi, \delta, \mu_\chi)$ parameter space, using a quenching factor $q_I = 0.08$.   We have neglected scattering off Na, as this is only important for $\delta \lesssim 30$ keV.  For the dark matter velocity distribution, we take a Maxwell-Boltzmann distribution with parameters $v_0 = 220$ km/s and $v_{esc} = 550$ km/s. We find the global $\chi^2$ minimum  and plot slices of allowed parameter space at constant $m_\chi$, showing the 90\% and 99\% confidence regions ($\Delta \chi^2 = 6.25$ and $11.34$, respectively), after ensuring that the best fit point gives a good fit to the data (i.e., $\chi^2/dof \lesssim 1$).  We note that higher mass dark matter are disfavored in fits that include the higher energy bins. The high recoil energy region strongly depends on the behavior of the form-factor and the two rows therefore serve as an illustration of the effects of form-factor uncertainties. We discuss the nuclear form factor uncertainties further below. We see that the scenario of inelastic dipole-dipole scattering is essentially unconstrained for a wide range of the inelasticity $\delta$.  The allowed range of $\mu_\chi \sim \rm{few} \times 10^{-3} \mu_N$ is comparable to what could arise from an electroweak loop, for instance, or if the WIMP is composed of charged constituents, bound by a force confining at the TeV scale, such as a technibaryon \cite{Bagnasco:1993st}.  

For our constraints, we use the full CDMS dataset. For ZEPLIN-III \cite{Akimov:2010vk} and XENON10 \cite{Collaboration:2009xb} we use recently published reanalyses focused on inelastic dark matter. The experiments closest to constraining the allowed DAMA region in our scenario are KIMS and ZEPLIN-III; ZEPLIN-III and CRESST have weakened limits relative to standard iDM \cite{Kuhlen:2009vh}, due to their suppressed magnetic couplings.  For Zeplin-III, a 90\% CL limit of 5.4 signal events \cite{Akimov:2010vk} was extracted using the maximum patch technique.  However, it is worth noting that the maximum patch technique used in \cite{Akimov:2010vk} would likely have different sensitivity for MiDM's altered recoil spectrum. Finally, we note that other lighter targets with large dipole moments, such as flourine and sodium, play essentially no role in the scenarios we consider. This is so because the inelastic threshold, $\delta$, is too large compared with the kinetic energy in the center of mass frame, $\mu_{\scriptstyle{N\chi}} v^2/2$, even for the highest speeds available in the halo.

Following the scattering against the nucleus, the excited DM state leaves the target and de-excites shortly after with a lifetime given by $\tau^{-1} = \mu_\chi^2 \delta^3/\pi$. For a typical splitting of $\delta = 120 \keV$ and a dipole moment $\mu_\chi = 3\times 10^{-3} \mu_{\scriptstyle N}$ the lifetime is $\tau = 5.1\; \mu s$, leading to an average decay length of $v \tau \sim 1.5$ m. This can give de-excitations inside the detector for a fraction of the events depositing a photon of energy $\delta \sim 100$ keV.  We note that in more complicated scenarios with multiple states, the de-excitation may occur to a different state other than the ground state, so the photon energy needn't be precisely equal to that of the excitation energy. (A similar phenomenology can occur in resonant dark matter (rDM), although with a photon of typically much higher energy, $\sim$ 500 keV -1 MeV \cite{Bai:2009cd}.)

Thus, the decay length of the excited state can be within an order of magnitude of the size of modern detectors.  Consequently, additional light could be added to WIMP scatters, making them appear either anomalous or more similar to electron recoils. In particular, in a large xenon detector (such as XENON100 or LUX) such an effect could potentially push the WIMP signal up, out of the nuclear recoil band, while leaving a signal at small detectors. Such decays could either give a signal of this scenario (if a second population appears, in addition to the primary one), or remove most of the nuclear recoil band signal, if the majority of the decays occur inside the detector. At DAMA in particular, the majority of the decays would occur either outside the experiment or inside a different crystal from the one where the initial scatter occured. Events separated by $>600$ ns are not counted in the multi-hit rate, and consequently existing cuts will \textit{not} veto against these events.  This higher energy modulation signal at $\delta$, due to the de-excitation,  is consistent with their observation of no modulation above 90 keVee \cite{Bernabei:2010mq}, but could be searched for with finer binning.  Other experiments, however, and in particular, xenon experiments with large volumes might reject these events by vetoes designed to remove multiple scatter events since the de-excitations occur displaced with respect to the initial scatter and yield a distorted signal shape. These de-excitations are a potentially exciting and useful signature, and we defer additional discussion to future work.

%Subsection
\subsection{Dark Dipole Scattering}
While a sizeable dipole of standard magnetism is possible, the WIMP could instead have a large {\em dark} magnetic dipole under some additional $U(1)$ which is kinetically mixed with electromagnetism through $\epsilon F_{\mu\nu}F_{\scriptstyle{D}}^{\mu\nu}/2$~\cite{Holdom:1985ag}. Such a feature would be particularly natural in composite inelastic models (CiDM) \cite{Alves:2009nf,Lisanti:2009am}. While the interaction is suppressed by the mixing parameter $\epsilon$, it can be enhanced in that the constraints on new dark charged constituents are far weaker than those for ordinary electromagnetism.

\begin{figure*}
\includegraphics[width=5.cm]{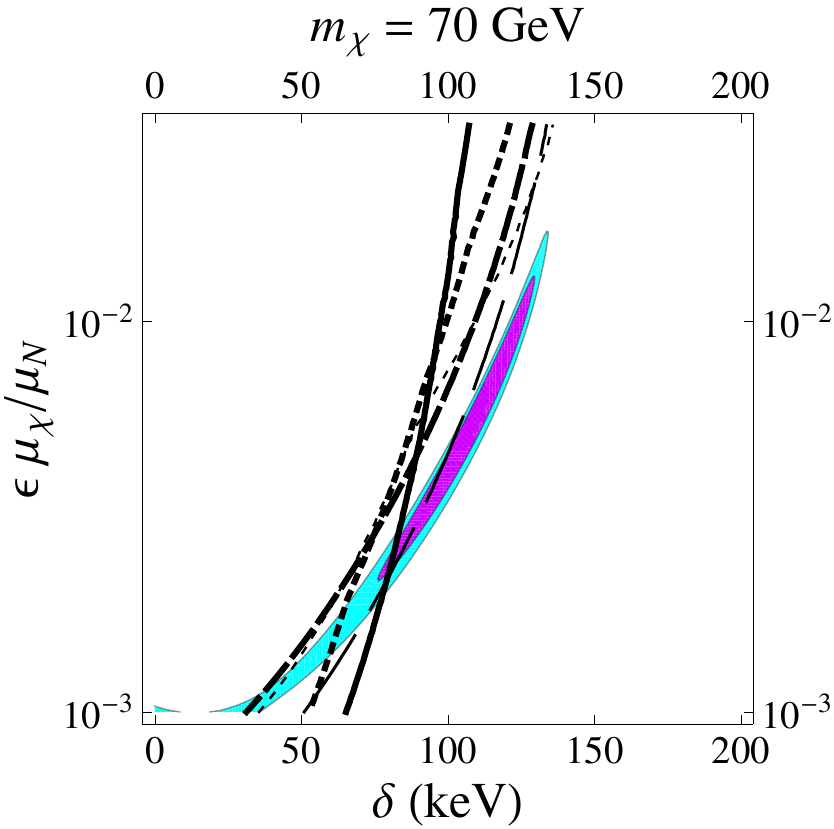}
\includegraphics[width=5.cm]{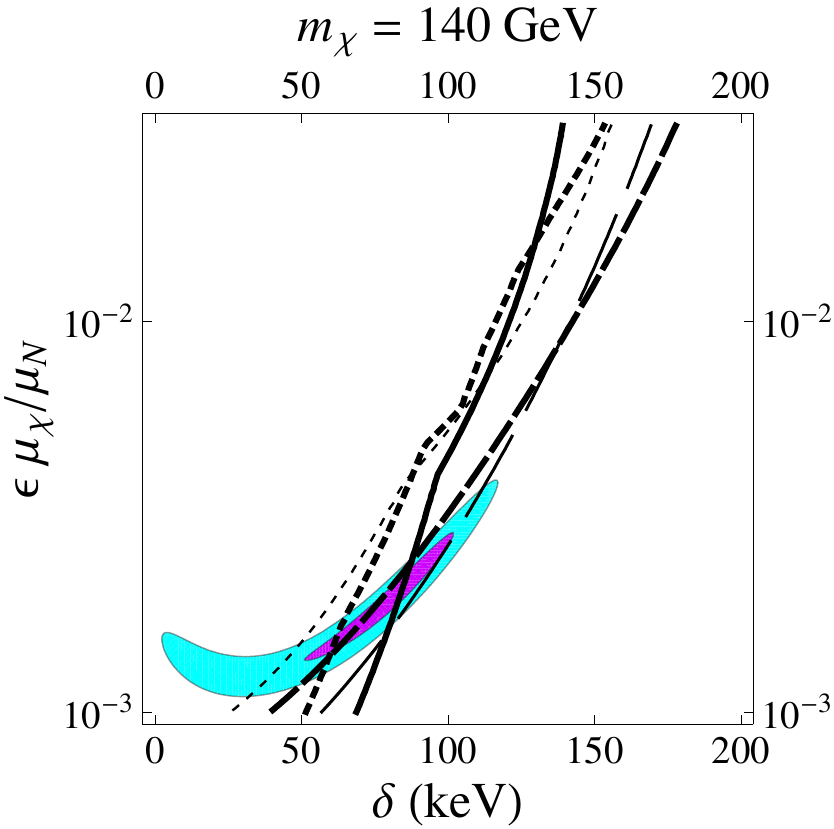}
\includegraphics[width=5.cm]{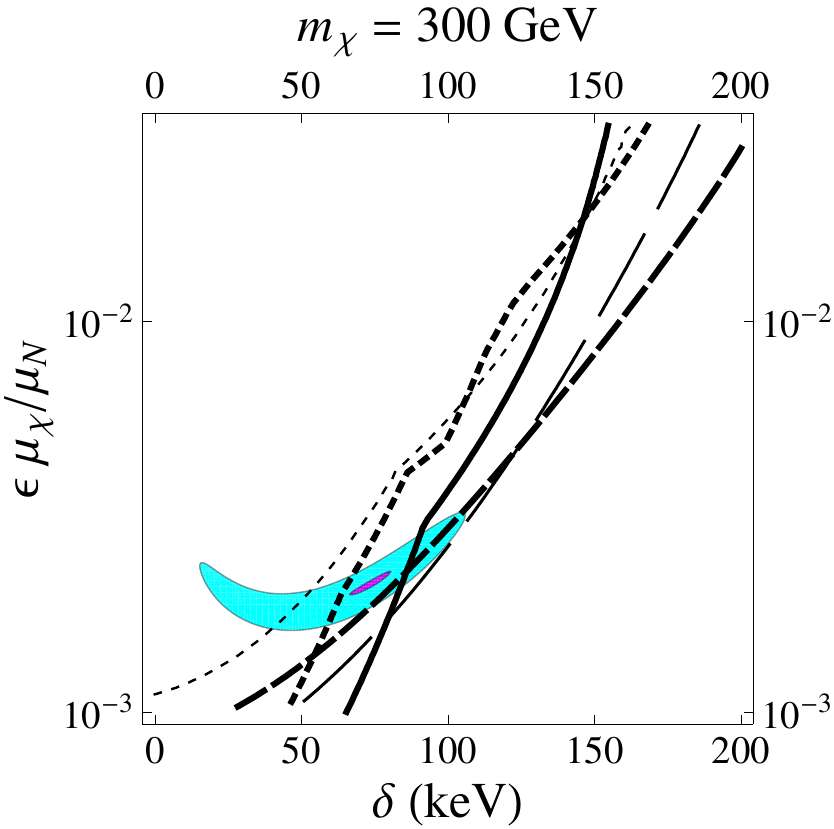}\\
\includegraphics[width=5.cm]{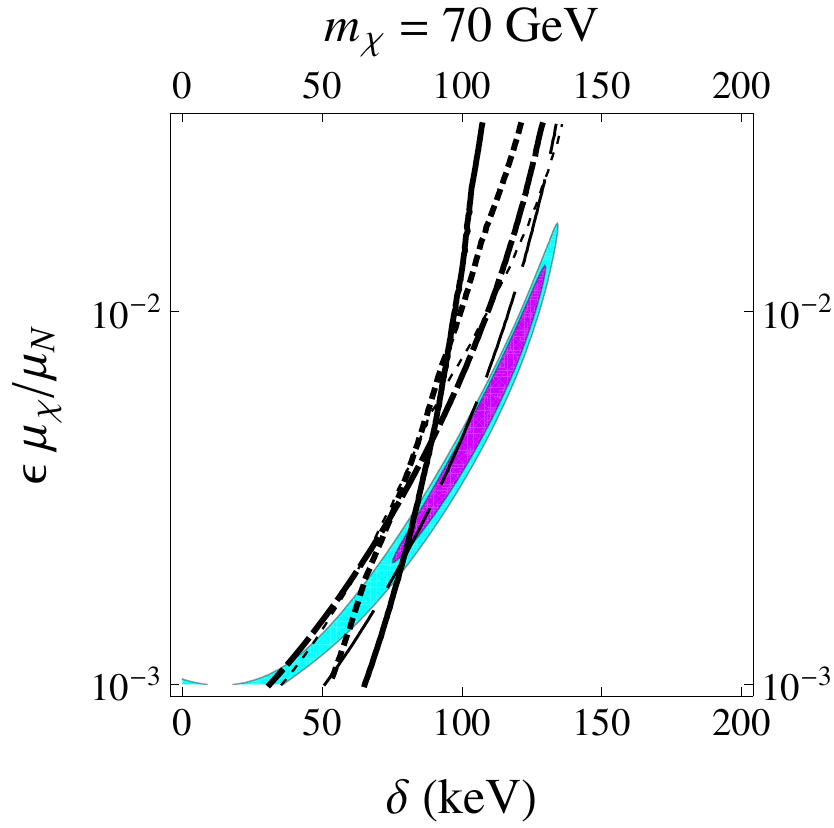}
\includegraphics[width=5.cm]{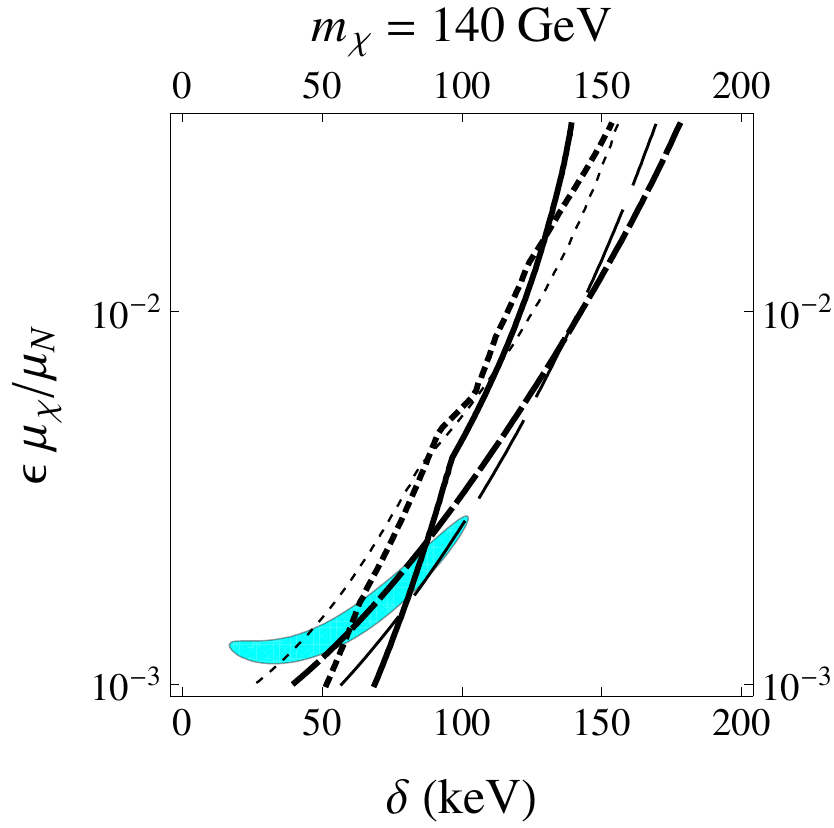}
\includegraphics[width=5.cm]{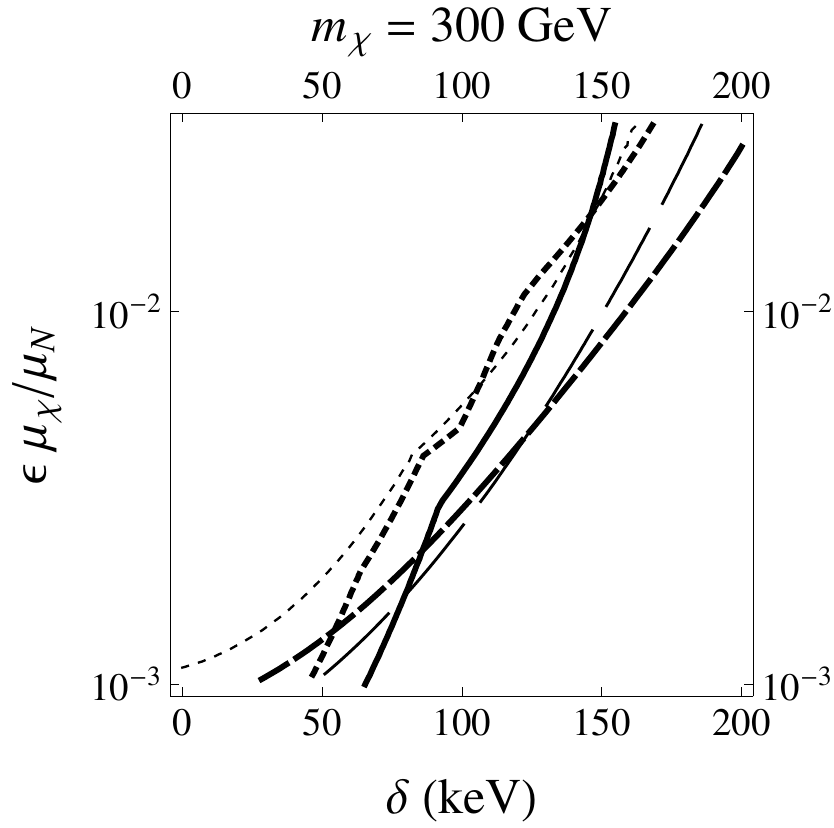}
\caption{Allowed ranges of parameter space for dark dipole scattering with mediator mass $m_{\darkmed} = 70\MeV$.  In this plot, we have set the dark gauge coupling $\alpha_d = \alpha$.  Lines are labeled as in Fig.~\ref{fig:dipoledipoleregions}. In the upper row, we utilized only the $2-8~$keVee bins in DAMA, and in the lower row we used the entire range of $2-14~$keVee. This serves to illustrate the strong dependence of the allowed DAMA parameter space for heavy dark matter on the dipole form factor's behavior at high $E_R$ as discussed in Sec.~\ref{sec:formfactor} in detail.}
\label{fig:darkdipoledipoleregions}
\end{figure*}

In the limit that the mediator is very light $m_{\darkmed}^2 \ll q^2 \equiv 2 m_N E_R$, the parameter space in Fig.~\ref{fig:dipoledipoleregions} is unchanged, but with $\mu_\chi \rightarrow \mu_\chi \epsilon$. We can also consider $m_{\darkmed}^2 \gg q^2$. In this case, the cross section is a simple modification of the QED formula, Eq.~(\ref{eqn:QEDformula}),  
\be
\frac{d \sigma_{dark}}{d \ER} = \frac{\epsilon^2 q^4}{(q^2+m_{\darkmed}^2)^2}\frac{d \sigma}{d \ER}
\ee
For the case where $m_{\darkmed} = 70 \MeV$, the parameter space is shown in Fig.~\ref{fig:darkdipoledipoleregions}. In this scenario, neglecting Na scattering at DAMA is an excellent approximation, as sodium is additionally suppressed relative to iodine by the $q^2$ factors. An interesting effect from the additional $q^4$ in the numerator is that the DAMA fit improves for smaller values of $\delta$, with a closer fit to the modulation shape. Larger values of the mediator mass, $m_{\darkmed}$, result in an increase (decrease) in the allowed parameter space shown in Fig.~\ref{fig:darkdipoledipoleregions} for small (large) WIMP masses, $\mX \lesssim 70\GeV$ ($\mX \gtrsim 70\GeV$). The cross sections required to fit DAMA increase for a heavier mediator, with a shift to lower values of $\delta$.

\subsection{Form Factor Uncertainties \label{sec:formfactor}}
As is well-known, spin-independent elastic scattering is suppressed at higher energies by the nuclear form factor. Explicit measurements of $F^2[\ER]$ have been made (see the discussion in \cite{Duda:2006uk}), and are in good agreement with the Helm form factor, which we employ for the contributions from dipole-charge scattering.

In contrast, there are no explicit measurements of the magnetic dipole form factor, or well-developed calculations available in the literature. The magnetic dipole moment arises from a variety of contributions in different nuclei, with a significant contribution often from proton angular momentum (in iodine, in particular) as well as from proton and neutron spin \cite{Ressell:1997kx}. Odd group and single particle models give soft form factors for the magnetic moment \cite{Ellis:1991ef}. \cite{Bagnasco:1993st} employs an estimated ``spin radius'', which gives a similarly soft form factor. However, these approaches neglect the nuclear spin contribution, whose form factors tend to have a hard tail, and the contribution of this should be taken into account.

Thus, we employ an approximation for the magnetic form factor: we take the contributions of angular momentum for iodine and cesium to be arising from, respectively, a 2d-shell orbital and a 1g-shell orbital, as expected in the nuclear shell model (see for e.g. \cite{Ellis:1991ef}), which we calculate explicitly using a SHO model of the nucleus.  For the SHO, we use a harmonic oscillator parameter of $b=2.282$ fm, which characterizes the size for an iodine nucleus \cite{Ressell:1997kx}.  We add to this a contribution from nuclear spin, which is taken from \cite{Ressell:1997kx} (I) and \cite{Toivanen:2009zza} (Cs). Since only squared-form factors are available, we take a square root and assume no sign change takes place in order to combine with the angular momentum contribution. We weight the angular momentum and spin contributions to the magnetic dipole moment as suggested by the values in \cite{Ressell:1997kx, Toivanen:2009zza}.  Specifically, we use
\bea
\nonumber
{\rm I}&:&  F_D^2[\ER]=\left(0.5 \frac{L_{2d}(E_R)}{L_{2d}(0)}+ 0.5 \sqrt{\frac{S_{pp}(E_R)}{S_{pp}(0)}}\right)^2\\[-.2cm] \\[-.05cm]
\nonumber
{\rm Cs}&:& F_D^2[\ER]=\left(1.35 \frac{L_{1g}(E_R)}{L_{1g}(0)}-0.35\sqrt{\frac{S_{pp}(E_R)}{S_{pp}(0)}}\right)^2
\eea
which characterize the contributions of the proton spin and angular momentum to the magnetic dipole moment for iodine and cesium which are both proton-odd.   The predicted scattering rate will depend on these coefficients in the combination, but of these the largest dependence is in cesium due to its destructive interference.  We have chosen the more conservative result that comes from using renormalized $g$ factors \cite{Toivanen:2009zza}.  It is worth noting that only after using these renormalized factors, is the magnetic dipole moment of Cs accurately predicted in \cite{Toivanen:2009zza}.  Using the unrenormalized factors would lead to a stronger KIMS limit, especially at higher WIMP masses.
The other heavy targets we consider have dipoles dominated by neutron spin, so we use the standard neutron spin form factor  for their scattering.

We should emphasize that combining the spin and angular momentum contributions in this way is conservative. Softer form factors tend to give better agreement for higher masses, and the destructive interference we assume for Cs means that the overall magnetic dipole form factor falls off slowly with $E_R$, since the cancellation ceases to be effective as the angular momentum contribution drops off rapidly.   This results in particularly stringent limits from KIMS. 
Given these uncertainties, we have employed an overall event rate constraint from KIMS, requiring the event rate from $3-8$ keVee (equivalently $20-100\; {\rm keV_R}$) to be less than the $90\%$ confidence level region of the observed rate $.28 \pm .16$ cpd/kg.  Note that this energy recoil range roughly corresponds to the $2-8$ keVee region we analyze for DAMA. 
On the other hand, since signals from Xe and W are dominated by dipole-charge scattering, their limits should be relatively robust.  

We emphasize that our choice of a DAMA $\chi^2$ fit to the 2-8 keVee bins shown in the first row of Figs.~\ref{fig:dipoledipoleregions} and \ref{fig:darkdipoledipoleregions} is insensitive to the form factor at $E_R \gtrsim 100\; {\rm keV_R}$. This row effectively illustrates the parameter space in case of a softer form-factor, since the  $8-14$ keVee bins would be less important for such a form factor. If the higher energy bins are included, as shown in the lower row of each figure, then a harder form-factor tail, such as the spin form factor, results in a poorer fit to DAMA for heavy dark matter masses (and in particular at large $\delta$). For example, we can see in Fig.~\ref{fig:dipoledipoleregions} and \ref{fig:darkdipoledipoleregions}, how the DAMA region shrinks entirely when the higher energy bins are included for the $\mX = 300$ GeV plots, while the plots for $\mX = 70\GeV$ are nearly unaffected.  Thus, whether or not heavy WIMPs are allowed in this scenario is very sensitive to the high energy behavior of the dipole form factor.  Consequently, numerical calculations of the form factor would help to pin down this parameter space.  Note that in the cases where the WIMP is composite, additional form factors are possible which may soften the spectrum at high energy, adding additional uncertainty beyond those of the nuclear magnetic dipole form factor.

\section{Conclusions}

The iodine target used at DAMA is set apart from other direct detection targets by two intrinsic properties:  its large mass and nuclear magnetic moment. Therefore, inelastic magnetic transitions which naturally utilize significant dipole-dipole interactions in addition to dipole-charge can explain the discrepancy between DAMA and other searches. In this letter, we considered this possibility in detail for WIMPs with magnetic dipoles or alternatively dark magnetic dipoles. We find that large parts of parameter space are available where a good fit to the DAMA modulation data is possible without violating constraints imposed by other searches. Given this largely untested scenario, we strongly encourage calculations of the nuclear magnetic dipole form factor, as this remaining uncertainty is important to determine the parameter ranges of interest.

{\it Magnetic}-like interactions also change our expectations for upcoming experiments.  
When the scattering at DAMA is enhanced by a sizeable dipole-dipole contribution, the signal experiments with small magnetic dipoles is suppressed by a comparable amount. At any of the xenon experiments, CRESST or CDMS, this is an additional suppression of a factor of a few. Moreover, as the lifetime of the excited state can be short, decays of the excited state in the detector can mask it as a double scattering event, potentially suppressing signal even farther in large xenon experiments. Thus, compared to standard iDM, a smaller signal would be expected in the current run at XENON100. On the other hand, KIMS with its CsI target retains sensitivity to this scenario and could search for it by either repeating their analysis with additional exposure or looking for modulation.  Of course, should the WIMP have both electric and magnetic dipoles mediating the inelastic transitions, then intermediate scenarios are possible. However, in all cases it seems that with $\sim$ 4000 kg day of summer exposure, XENON100 should be able to definitively test any scenario in which scattering of the magnetic dipole plays a major role. Because de-excitation can occur within the detector volume, standard cuts may miss a sizable fraction of these events. However, if attention is paid to double-scatter coincident events, xenon experiments should be capable of detecting both the inelastic WIMP signal and its magnetic origin.

In conclusion, we have considered models of iDM where dark matter has a (EM or dark) magnetic dipole, which plays a major role in its scattering off of nuclei. The presence of the large magnetic dipole moment of iodine in addition to its large mass widens the parameter space where the modulation signal at DAMA is consistent with the null results from all other experiments.   
All models predict some signal at the XENON100, KIMS, and CRESST experiments, with the usual iDM spectral features, and a large modulation amplitude. With new data on the horizon, these models should be tested soon.

 \begin{acknowledgments}
We would like to thank Laura Baudis, Pierluigi Belli, Rafael Lang, Dan McKinsey, and Peter Sorensen for useful discussions pertaining to the different experiments. We are grateful to  Wick Haxton and David Dean for their insightful and extremely patient discussions of the spin structure of nucleons. We thank T.Banks and S. Thomas for useful discussions. They have informed us that they are writing a paper with J-F. Fortin on the signals for dark matter with electromagnetic dipoles in terrestrial detectors.  While there is some overlap with our work (they also emphasize the importance of the large magnetic moment of Iodine), their models do not have inelastic scattering and differ significantly from our own. SC is supported under DOE Grant \#DE-FG02-91ER40674. NW is supported by DOE OJI grant \#DE-FG02-06ER41417 and NSF grant \#0947827. IY is supported by the James Arthur fellowship. 
\end{acknowledgments}
\bibliography{DMDM}

%merlin.mbs 2010-03-15 4.21a (PWD, AO, DPC)
\begin{thebibliography}{68}%
\makeatletter
\providecommand \@ifxundefined [1]{%
 \@ifx{#1\undefined}
}%
\providecommand \@ifnum [1]{%
 \ifnum #1\expandafter \@firstoftwo
 \else \expandafter \@secondoftwo
 \fi
}%
\providecommand \@ifx [1]{%
 \ifx #1\expandafter \@firstoftwo
 \else \expandafter \@secondoftwo
 \fi
}%
\providecommand \natexlab [1]{#1}%
\providecommand \enquote  [1]{``#1''}%
\providecommand \bibnamefont  [1]{#1}%
\providecommand \bibfnamefont [1]{#1}%
\providecommand \citenamefont [1]{#1}%
\providecommand \href@noop [0]{\@secondoftwo}%
\providecommand \href [0]{\begingroup \@sanitize@url \@href}%
\providecommand \@href[1]{\@@startlink{#1}\@@href}%
\providecommand \@@href[1]{\endgroup#1\@@endlink}%
\providecommand \@sanitize@url [0]{\catcode `\\12\catcode `\$12\catcode
  `\&12\catcode `\#12\catcode `\^12\catcode `\_12\catcode `\%12\relax}%
\providecommand \@@startlink[1]{}%
\providecommand \@@endlink[0]{}%
\providecommand \url  [0]{\begingroup\@sanitize@url \@url }%
\providecommand \@url [1]{\endgroup\@href {#1}{\urlprefix }}%
\providecommand \urlprefix  [0]{URL }%
\providecommand \Eprint [0]{\href }%
\@ifxundefined \urlstyle {%
  \providecommand \doi  [0]{\begingroup \@sanitize@url \@doi}%
  \providecommand \@doi [1]{\endgroup \@@startlink {\doibase
  #1}doi:\discretionary {}{}{}#1\@@endlink }%
}{%
  \providecommand \doi  [0]{doi:\discretionary{}{}{}\begingroup
  \urlstyle{rm}\Url }%
}%
\providecommand \doibase [0]{http://dx.doi.org/}%
\providecommand \Doi [0]{\begingroup \@sanitize@url \@Doi }%
\providecommand \@Doi  [1]{\endgroup\@@startlink{\doibase#1}\@@Doi}%
\providecommand \@@Doi [1]{#1\@@endlink}%
\providecommand \selectlanguage [0]{\@gobble}%
\providecommand \bibinfo  [0]{\@secondoftwo}%
\providecommand \bibfield  [0]{\@secondoftwo}%
\providecommand \translation [1]{[#1]}%
\providecommand \BibitemOpen [0]{}%
\providecommand \bibitemStop [0]{}%
\providecommand \bibitemNoStop [0]{.\EOS\space}%
\providecommand \EOS [0]{\spacefactor3000\relax}%
\providecommand \BibitemShut  [1]{\csname bibitem#1\endcsname}%
%</preamble>
\bibitem [{\citenamefont {Bernabei}\ \emph {et~al.}(2010)\citenamefont
  {Bernabei} \emph {et~al.}}]{Bernabei:2010mq}%
  \BibitemOpen
  \bibfield  {author} {\bibinfo {author} {\bibfnamefont {R.}~\bibnamefont
  {Bernabei}} \emph {et~al.},\ }\Doi {10.1140/epjc/s10052-010-1303-9} {
  (\bibinfo {year} {2010})},\ \doi {10.1140/epjc/s10052-010-1303-9},\ \Eprint
  {http://arxiv.org/abs/1002.1028} {arXiv:1002.1028 [astro-ph.GA]} \BibitemShut
  {NoStop}%
%%CITATION = 1002.1028;%%
\bibitem [{\citenamefont {Aprile}\ \emph {et~al.}(2010)\citenamefont {Aprile}
  \emph {et~al.}}]{Aprile:2010um}%
  \BibitemOpen
  \bibfield  {author} {\bibinfo {author} {\bibfnamefont {E.}~\bibnamefont
  {Aprile}} \emph {et~al.} (\bibinfo {collaboration} {XENON100}),\ }\href@noop
  {} { (\bibinfo {year} {2010})},\ \Eprint {http://arxiv.org/abs/1005.0380}
  {arXiv:1005.0380 [astro-ph.CO]} \BibitemShut {NoStop}%
%%CITATION = 1005.0380;%%
\bibitem [{\citenamefont {Ahmed}\ \emph {et~al.}(2009)\citenamefont {Ahmed}
  \emph {et~al.}}]{Ahmed:2009zw}%
  \BibitemOpen
  \bibfield  {author} {\bibinfo {author} {\bibfnamefont {Z.}~\bibnamefont
  {Ahmed}} \emph {et~al.} (\bibinfo {collaboration} {The CDMS-II}),\
  }\href@noop {} { (\bibinfo {year} {2009})},\ \Eprint
  {http://arxiv.org/abs/0912.3592} {arXiv:0912.3592 [astro-ph.CO]} \BibitemShut
  {NoStop}%
%%CITATION = 0912.3592;%%
\bibitem [{\citenamefont {Sanglard}(2010)}]{Sanglard:2009qp}%
  \BibitemOpen
  \bibfield  {author} {\bibinfo {author} {\bibfnamefont {V.}~\bibnamefont
  {Sanglard}} (\bibinfo {collaboration} {for the EDELWEISS}),\ }\Doi
  {10.1088/1742-6596/203/1/012037} {\bibfield  {journal} {\bibinfo  {journal}
  {J. Phys. Conf. Ser.},\ }\textbf {\bibinfo {volume} {203}},\ \bibinfo {pages}
  {012037} (\bibinfo {year} {2010})},\ \Eprint {http://arxiv.org/abs/0912.1196}
  {arXiv:0912.1196 [astro-ph.CO]} \BibitemShut {NoStop}%
%%CITATION = 0912.1196;%%
\bibitem [{\citenamefont {Lang}\ and\ \citenamefont
  {Seidel}(2009)}]{Lang:2009ge}%
  \BibitemOpen
  \bibfield  {author} {\bibinfo {author} {\bibfnamefont {R.~F.}\ \bibnamefont
  {Lang}}\ and\ \bibinfo {author} {\bibfnamefont {W.}~\bibnamefont {Seidel}},\
  }\Doi {10.1088/1367-2630/11/10/105017} {\bibfield  {journal} {\bibinfo
  {journal} {New J. Phys.},\ }\textbf {\bibinfo {volume} {11}},\ \bibinfo
  {pages} {105017} (\bibinfo {year} {2009})},\ \Eprint
  {http://arxiv.org/abs/0906.3290} {arXiv:0906.3290 [astro-ph.IM]} \BibitemShut
  {NoStop}%
%%CITATION = 0906.3290;%%
\bibitem [{\citenamefont {Lebedenko}\ \emph {et~al.}(2009)\citenamefont
  {Lebedenko} \emph {et~al.}}]{Lebedenko:2009xe}%
  \BibitemOpen
  \bibfield  {author} {\bibinfo {author} {\bibfnamefont {V.~N.}\ \bibnamefont
  {Lebedenko}} \emph {et~al.} (\bibinfo {collaboration} {ZEPLIN-III}),\ }\Doi
  {10.1103/PhysRevLett.103.151302} {\bibfield  {journal} {\bibinfo  {journal}
  {Phys. Rev. Lett.},\ }\textbf {\bibinfo {volume} {103}},\ \bibinfo {pages}
  {151302} (\bibinfo {year} {2009})},\ \Eprint {http://arxiv.org/abs/0901.4348}
  {arXiv:0901.4348 [hep-ex]} \BibitemShut {NoStop}%
%%CITATION = 0901.4348;%%
\bibitem [{\citenamefont {Ullio}\ \emph {et~al.}(2001)\citenamefont {Ullio},
  \citenamefont {Kamionkowski},\ and\ \citenamefont {Vogel}}]{Ullio:2000bv}%
  \BibitemOpen
  \bibfield  {author} {\bibinfo {author} {\bibfnamefont {P.}~\bibnamefont
  {Ullio}}, \bibinfo {author} {\bibfnamefont {M.}~\bibnamefont {Kamionkowski}},
  \ and\ \bibinfo {author} {\bibfnamefont {P.}~\bibnamefont {Vogel}},\
  }\href@noop {} {\bibfield  {journal} {\bibinfo  {journal} {JHEP},\ }\textbf
  {\bibinfo {volume} {07}},\ \bibinfo {pages} {044} (\bibinfo {year} {2001})},\
  \Eprint {http://arxiv.org/abs/hep-ph/0010036} {arXiv:hep-ph/0010036}
  \BibitemShut {NoStop}%
%%CITATION = HEP-PH/0010036;%%
\bibitem [{\citenamefont {Belli}\ \emph {et~al.}(2002)\citenamefont {Belli},
  \citenamefont {Cerulli}, \citenamefont {Fornengo},\ and\ \citenamefont
  {Scopel}}]{Belli:2002yt}%
  \BibitemOpen
  \bibfield  {author} {\bibinfo {author} {\bibfnamefont {P.}~\bibnamefont
  {Belli}}, \bibinfo {author} {\bibfnamefont {R.}~\bibnamefont {Cerulli}},
  \bibinfo {author} {\bibfnamefont {N.}~\bibnamefont {Fornengo}}, \ and\
  \bibinfo {author} {\bibfnamefont {S.}~\bibnamefont {Scopel}},\ }\Doi
  {10.1103/PhysRevD.66.043503} {\bibfield  {journal} {\bibinfo  {journal}
  {Phys. Rev.},\ }\textbf {\bibinfo {volume} {D66}},\ \bibinfo {pages} {043503}
  (\bibinfo {year} {2002})},\ \Eprint {http://arxiv.org/abs/hep-ph/0203242}
  {arXiv:hep-ph/0203242} \BibitemShut {NoStop}%
%%CITATION = HEP-PH/0203242;%%
\bibitem [{\citenamefont {Savage}\ \emph {et~al.}(2004)\citenamefont {Savage},
  \citenamefont {Gondolo},\ and\ \citenamefont {Freese}}]{Savage:2004fn}%
  \BibitemOpen
  \bibfield  {author} {\bibinfo {author} {\bibfnamefont {C.}~\bibnamefont
  {Savage}}, \bibinfo {author} {\bibfnamefont {P.}~\bibnamefont {Gondolo}}, \
  and\ \bibinfo {author} {\bibfnamefont {K.}~\bibnamefont {Freese}},\ }\Doi
  {10.1103/PhysRevD.70.123513} {\bibfield  {journal} {\bibinfo  {journal}
  {Phys. Rev.},\ }\textbf {\bibinfo {volume} {D70}},\ \bibinfo {pages} {123513}
  (\bibinfo {year} {2004})},\ \Eprint {http://arxiv.org/abs/astro-ph/0408346}
  {arXiv:astro-ph/0408346} \BibitemShut {NoStop}%
%%CITATION = ASTRO-PH/0408346;%%
\bibitem [{\citenamefont {Bernabei}\ \emph {et~al.}(2004)\citenamefont
  {Bernabei} \emph {et~al.}}]{Bernabei:2005hj}%
  \BibitemOpen
  \bibfield  {author} {\bibinfo {author} {\bibfnamefont {R.}~\bibnamefont
  {Bernabei}} \emph {et~al.},\ }\Doi {10.1142/S0218271804006619} {\bibfield
  {journal} {\bibinfo  {journal} {Int. J. Mod. Phys.},\ }\textbf {\bibinfo
  {volume} {D13}},\ \bibinfo {pages} {2127} (\bibinfo {year} {2004})},\ \Eprint
  {http://arxiv.org/abs/astro-ph/0501412} {arXiv:astro-ph/0501412} \BibitemShut
  {NoStop}%
%%CITATION = ASTRO-PH/0501412;%%
\bibitem [{\citenamefont {Fairbairn}\ and\ \citenamefont
  {Schwetz}(2009)}]{Fairbairn:2008gz}%
  \BibitemOpen
  \bibfield  {author} {\bibinfo {author} {\bibfnamefont {M.}~\bibnamefont
  {Fairbairn}}\ and\ \bibinfo {author} {\bibfnamefont {T.}~\bibnamefont
  {Schwetz}},\ }\Doi {10.1088/1475-7516/2009/01/037} {\bibfield  {journal}
  {\bibinfo  {journal} {JCAP},\ }\textbf {\bibinfo {volume} {0901}},\ \bibinfo
  {pages} {037} (\bibinfo {year} {2009})},\ \Eprint
  {http://arxiv.org/abs/0808.0704} {arXiv:0808.0704 [hep-ph]} \BibitemShut
  {NoStop}%
%%CITATION = 0808.0704;%%
\bibitem [{\citenamefont {Foot}(2004)}]{Foot:2004gh}%
  \BibitemOpen
  \bibfield  {author} {\bibinfo {author} {\bibfnamefont {R.}~\bibnamefont
  {Foot}},\ }\Doi {10.1142/S0217732304015051} {\bibfield  {journal} {\bibinfo
  {journal} {Mod. Phys. Lett.},\ }\textbf {\bibinfo {volume} {A19}},\ \bibinfo
  {pages} {1841} (\bibinfo {year} {2004})},\ \Eprint
  {http://arxiv.org/abs/astro-ph/0405362} {arXiv:astro-ph/0405362} \BibitemShut
  {NoStop}%
%%CITATION = ASTRO-PH/0405362;%%
\bibitem [{\citenamefont {Foot}(2008)}]{Foot:2008nw}%
  \BibitemOpen
  \bibfield  {author} {\bibinfo {author} {\bibfnamefont {R.}~\bibnamefont
  {Foot}},\ }\Doi {10.1103/PhysRevD.78.043529} {\bibfield  {journal} {\bibinfo
  {journal} {Phys. Rev.},\ }\textbf {\bibinfo {volume} {D78}},\ \bibinfo
  {pages} {043529} (\bibinfo {year} {2008})},\ \Eprint
  {http://arxiv.org/abs/0804.4518} {arXiv:0804.4518 [hep-ph]} \BibitemShut
  {NoStop}%
%%CITATION = 0804.4518;%%
\bibitem [{\citenamefont {Bai}\ and\ \citenamefont {Fox}(2009)}]{Bai:2009cd}%
  \BibitemOpen
  \bibfield  {author} {\bibinfo {author} {\bibfnamefont {Y.}~\bibnamefont
  {Bai}}\ and\ \bibinfo {author} {\bibfnamefont {P.~J.}\ \bibnamefont {Fox}},\
  }\Doi {10.1088/1126-6708/2009/11/052} {\bibfield  {journal} {\bibinfo
  {journal} {JHEP},\ }\textbf {\bibinfo {volume} {11}},\ \bibinfo {pages} {052}
  (\bibinfo {year} {2009})},\ \Eprint {http://arxiv.org/abs/0909.2900}
  {arXiv:0909.2900 [hep-ph]} \BibitemShut {NoStop}%
%%CITATION = 0909.2900;%%
\bibitem [{\citenamefont {Essig}\ \emph {et~al.}(2010)\citenamefont {Essig},
  \citenamefont {Kaplan}, \citenamefont {Schuster},\ and\ \citenamefont
  {Toro}}]{Essig:2010ye}%
  \BibitemOpen
  \bibfield  {author} {\bibinfo {author} {\bibfnamefont {R.}~\bibnamefont
  {Essig}}, \bibinfo {author} {\bibfnamefont {J.}~\bibnamefont {Kaplan}},
  \bibinfo {author} {\bibfnamefont {P.}~\bibnamefont {Schuster}}, \ and\
  \bibinfo {author} {\bibfnamefont {N.}~\bibnamefont {Toro}},\ }\href@noop {} {
  (\bibinfo {year} {2010})},\ \Eprint {http://arxiv.org/abs/1004.0691}
  {arXiv:1004.0691 [hep-ph]} \BibitemShut {NoStop}%
%%CITATION = 1004.0691;%%
\bibitem [{\citenamefont {Graham}\ \emph {et~al.}(2010)\citenamefont {Graham},
  \citenamefont {Harnik}, \citenamefont {Rajendran},\ and\ \citenamefont
  {Saraswat}}]{Graham:2010ca}%
  \BibitemOpen
  \bibfield  {author} {\bibinfo {author} {\bibfnamefont {P.~W.}\ \bibnamefont
  {Graham}}, \bibinfo {author} {\bibfnamefont {R.}~\bibnamefont {Harnik}},
  \bibinfo {author} {\bibfnamefont {S.}~\bibnamefont {Rajendran}}, \ and\
  \bibinfo {author} {\bibfnamefont {P.}~\bibnamefont {Saraswat}},\ }\href@noop
  {} { (\bibinfo {year} {2010})},\ \Eprint {http://arxiv.org/abs/1004.0937}
  {arXiv:1004.0937 [hep-ph]} \BibitemShut {NoStop}%
%%CITATION = 1004.0937;%%
\bibitem [{\citenamefont {Feldstein}\ \emph {et~al.}(2010)\citenamefont
  {Feldstein}, \citenamefont {Fitzpatrick},\ and\ \citenamefont
  {Katz}}]{Feldstein:2009tr}%
  \BibitemOpen
  \bibfield  {author} {\bibinfo {author} {\bibfnamefont {B.}~\bibnamefont
  {Feldstein}}, \bibinfo {author} {\bibfnamefont {A.~L.}\ \bibnamefont
  {Fitzpatrick}}, \ and\ \bibinfo {author} {\bibfnamefont {E.}~\bibnamefont
  {Katz}},\ }\Doi {10.1088/1475-7516/2010/01/020} {\bibfield  {journal}
  {\bibinfo  {journal} {JCAP},\ }\textbf {\bibinfo {volume} {1001}},\ \bibinfo
  {pages} {020} (\bibinfo {year} {2010})},\ \Eprint
  {http://arxiv.org/abs/0908.2991} {arXiv:0908.2991 [hep-ph]} \BibitemShut
  {NoStop}%
%%CITATION = 0908.2991;%%
\bibitem [{\citenamefont {Chang}\ \emph
  {et~al.}(2010){\natexlab{a}}\citenamefont {Chang}, \citenamefont {Pierce},\
  and\ \citenamefont {Weiner}}]{Chang:2009yt}%
  \BibitemOpen
  \bibfield  {author} {\bibinfo {author} {\bibfnamefont {S.}~\bibnamefont
  {Chang}}, \bibinfo {author} {\bibfnamefont {A.}~\bibnamefont {Pierce}}, \
  and\ \bibinfo {author} {\bibfnamefont {N.}~\bibnamefont {Weiner}},\ }\Doi
  {10.1088/1475-7516/2010/01/006} {\bibfield  {journal} {\bibinfo  {journal}
  {JCAP},\ }\textbf {\bibinfo {volume} {1001}},\ \bibinfo {pages} {006}
  (\bibinfo {year} {2010}{\natexlab{a}})},\ \Eprint
  {http://arxiv.org/abs/0908.3192} {arXiv:0908.3192 [hep-ph]} \BibitemShut
  {NoStop}%
%%CITATION = 0908.3192;%%
\bibitem [{\citenamefont {Belli}\ \emph {et~al.}(2000)\citenamefont {Belli}
  \emph {et~al.}}]{Belli:1999nz}%
  \BibitemOpen
  \bibfield  {author} {\bibinfo {author} {\bibfnamefont {P.}~\bibnamefont
  {Belli}} \emph {et~al.},\ }\Doi {10.1103/PhysRevD.61.023512} {\bibfield
  {journal} {\bibinfo  {journal} {Phys. Rev.},\ }\textbf {\bibinfo {volume}
  {D61}},\ \bibinfo {pages} {023512} (\bibinfo {year} {2000})},\ \Eprint
  {http://arxiv.org/abs/hep-ph/9903501} {arXiv:hep-ph/9903501} \BibitemShut
  {NoStop}%
%%CITATION = HEP-PH/9903501;%%
\bibitem [{\citenamefont {Bottino}\ \emph {et~al.}(2004)\citenamefont
  {Bottino}, \citenamefont {Donato}, \citenamefont {Fornengo},\ and\
  \citenamefont {Scopel}}]{Bottino:2003cz}%
  \BibitemOpen
  \bibfield  {author} {\bibinfo {author} {\bibfnamefont {A.}~\bibnamefont
  {Bottino}}, \bibinfo {author} {\bibfnamefont {F.}~\bibnamefont {Donato}},
  \bibinfo {author} {\bibfnamefont {N.}~\bibnamefont {Fornengo}}, \ and\
  \bibinfo {author} {\bibfnamefont {S.}~\bibnamefont {Scopel}},\ }\Doi
  {10.1103/PhysRevD.69.037302} {\bibfield  {journal} {\bibinfo  {journal}
  {Phys. Rev.},\ }\textbf {\bibinfo {volume} {D69}},\ \bibinfo {pages} {037302}
  (\bibinfo {year} {2004})},\ \Eprint {http://arxiv.org/abs/hep-ph/0307303}
  {arXiv:hep-ph/0307303} \BibitemShut {NoStop}%
%%CITATION = HEP-PH/0307303;%%
\bibitem [{\citenamefont {Bernabei}\ \emph {et~al.}(2008)\citenamefont
  {Bernabei} \emph {et~al.}}]{Bernabei:2007hw}%
  \BibitemOpen
  \bibfield  {author} {\bibinfo {author} {\bibfnamefont {R.}~\bibnamefont
  {Bernabei}} \emph {et~al.},\ }\Doi {10.1140/epjc/s10052-007-0479-0}
  {\bibfield  {journal} {\bibinfo  {journal} {Eur. Phys. J.},\ }\textbf
  {\bibinfo {volume} {C53}},\ \bibinfo {pages} {205} (\bibinfo {year}
  {2008})},\ \Eprint {http://arxiv.org/abs/0710.0288} {arXiv:0710.0288
  [astro-ph]} \BibitemShut {NoStop}%
%%CITATION = 0710.0288;%%
\bibitem [{\citenamefont {Bottino}\ \emph {et~al.}(2008)\citenamefont
  {Bottino}, \citenamefont {Donato}, \citenamefont {Fornengo},\ and\
  \citenamefont {Scopel}}]{Bottino:2008mf}%
  \BibitemOpen
  \bibfield  {author} {\bibinfo {author} {\bibfnamefont {A.}~\bibnamefont
  {Bottino}}, \bibinfo {author} {\bibfnamefont {F.}~\bibnamefont {Donato}},
  \bibinfo {author} {\bibfnamefont {N.}~\bibnamefont {Fornengo}}, \ and\
  \bibinfo {author} {\bibfnamefont {S.}~\bibnamefont {Scopel}},\ }\Doi
  {10.1103/PhysRevD.78.083520} {\bibfield  {journal} {\bibinfo  {journal}
  {Phys. Rev.},\ }\textbf {\bibinfo {volume} {D78}},\ \bibinfo {pages} {083520}
  (\bibinfo {year} {2008})},\ \Eprint {http://arxiv.org/abs/0806.4099}
  {arXiv:0806.4099 [hep-ph]} \BibitemShut {NoStop}%
%%CITATION = 0806.4099;%%
\bibitem [{\citenamefont {Chang}\ \emph
  {et~al.}(2008){\natexlab{a}}\citenamefont {Chang}, \citenamefont {Pierce},\
  and\ \citenamefont {Weiner}}]{Chang:2008xa}%
  \BibitemOpen
  \bibfield  {author} {\bibinfo {author} {\bibfnamefont {S.}~\bibnamefont
  {Chang}}, \bibinfo {author} {\bibfnamefont {A.}~\bibnamefont {Pierce}}, \
  and\ \bibinfo {author} {\bibfnamefont {N.}~\bibnamefont {Weiner}},\
  }\href@noop {} { (\bibinfo {year} {2008}{\natexlab{a}})},\ \Eprint
  {http://arxiv.org/abs/0808.0196} {arXiv:0808.0196 [hep-ph]} \BibitemShut
  {NoStop}%
%%CITATION = 0808.0196;%%
\bibitem [{\citenamefont {Savage}\ \emph {et~al.}(2009)\citenamefont {Savage},
  \citenamefont {Gelmini}, \citenamefont {Gondolo},\ and\ \citenamefont
  {Freese}}]{Savage:2008er}%
  \BibitemOpen
  \bibfield  {author} {\bibinfo {author} {\bibfnamefont {C.}~\bibnamefont
  {Savage}}, \bibinfo {author} {\bibfnamefont {G.}~\bibnamefont {Gelmini}},
  \bibinfo {author} {\bibfnamefont {P.}~\bibnamefont {Gondolo}}, \ and\
  \bibinfo {author} {\bibfnamefont {K.}~\bibnamefont {Freese}},\ }\Doi
  {10.1088/1475-7516/2009/04/010} {\bibfield  {journal} {\bibinfo  {journal}
  {JCAP},\ }\textbf {\bibinfo {volume} {0904}},\ \bibinfo {pages} {010}
  (\bibinfo {year} {2009})},\ \Eprint {http://arxiv.org/abs/0808.3607}
  {arXiv:0808.3607 [astro-ph]} \BibitemShut {NoStop}%
%%CITATION = 0808.3607;%%
\bibitem [{\citenamefont {Petriello}\ and\ \citenamefont
  {Zurek}(2008)}]{Petriello:2008jj}%
  \BibitemOpen
  \bibfield  {author} {\bibinfo {author} {\bibfnamefont {F.}~\bibnamefont
  {Petriello}}\ and\ \bibinfo {author} {\bibfnamefont {K.~M.}\ \bibnamefont
  {Zurek}},\ }\Doi {10.1088/1126-6708/2008/09/047} {\bibfield  {journal}
  {\bibinfo  {journal} {JHEP},\ }\textbf {\bibinfo {volume} {09}},\ \bibinfo
  {pages} {047} (\bibinfo {year} {2008})},\ \Eprint
  {http://arxiv.org/abs/0806.3989} {arXiv:0806.3989 [hep-ph]} \BibitemShut
  {NoStop}%
%%CITATION = 0806.3989;%%
\bibitem [{\citenamefont {Hooper}\ \emph {et~al.}(2010)\citenamefont {Hooper},
  \citenamefont {Collar}, \citenamefont {Hall},\ and\ \citenamefont
  {McKinsey}}]{Hooper:2010uy}%
  \BibitemOpen
  \bibfield  {author} {\bibinfo {author} {\bibfnamefont {D.}~\bibnamefont
  {Hooper}}, \bibinfo {author} {\bibfnamefont {J.~I.}\ \bibnamefont {Collar}},
  \bibinfo {author} {\bibfnamefont {J.}~\bibnamefont {Hall}}, \ and\ \bibinfo
  {author} {\bibfnamefont {D.}~\bibnamefont {McKinsey}},\ }\href@noop {} {
  (\bibinfo {year} {2010})},\ \Eprint {http://arxiv.org/abs/1007.1005}
  {arXiv:1007.1005 [hep-ph]} \BibitemShut {NoStop}%
%%CITATION = 1007.1005;%%
\bibitem [{\citenamefont {Tucker-Smith}\ and\ \citenamefont
  {Weiner}(2001)}]{TuckerSmith:2001hy}%
  \BibitemOpen
  \bibfield  {author} {\bibinfo {author} {\bibfnamefont {D.}~\bibnamefont
  {Tucker-Smith}}\ and\ \bibinfo {author} {\bibfnamefont {N.}~\bibnamefont
  {Weiner}},\ }\Doi {10.1103/PhysRevD.64.043502} {\bibfield  {journal}
  {\bibinfo  {journal} {Phys. Rev.},\ }\textbf {\bibinfo {volume} {D64}},\
  \bibinfo {pages} {043502} (\bibinfo {year} {2001})},\ \Eprint
  {http://arxiv.org/abs/hep-ph/0101138} {arXiv:hep-ph/0101138} \BibitemShut
  {NoStop}%
%%CITATION = HEP-PH/0101138;%%
\bibitem [{\citenamefont {Chang}\ \emph
  {et~al.}(2008){\natexlab{b}}\citenamefont {Chang}, \citenamefont {Kribs},
  \citenamefont {Tucker-Smith},\ and\ \citenamefont {Weiner}}]{WeinerKribs}%
  \BibitemOpen
  \bibfield  {author} {\bibinfo {author} {\bibfnamefont {S.}~\bibnamefont
  {Chang}}, \bibinfo {author} {\bibfnamefont {G.~D.}\ \bibnamefont {Kribs}},
  \bibinfo {author} {\bibfnamefont {D.}~\bibnamefont {Tucker-Smith}}, \ and\
  \bibinfo {author} {\bibfnamefont {N.}~\bibnamefont {Weiner}},\ }\href@noop {}
  { (\bibinfo {year} {2008}{\natexlab{b}})},\ \Eprint
  {http://arxiv.org/abs/0807.2250} {arXiv:0807.2250 [hep-ph]} \BibitemShut
  {NoStop}%
%%CITATION = 0807.2250;%%
\bibitem [{\citenamefont {March-Russell}\ \emph {et~al.}(2008)\citenamefont
  {March-Russell}, \citenamefont {McCabe},\ and\ \citenamefont
  {McCullough}}]{MarchRussell:2008dy}%
  \BibitemOpen
  \bibfield  {author} {\bibinfo {author} {\bibfnamefont {J.}~\bibnamefont
  {March-Russell}}, \bibinfo {author} {\bibfnamefont {C.}~\bibnamefont
  {McCabe}}, \ and\ \bibinfo {author} {\bibfnamefont {M.}~\bibnamefont
  {McCullough}},\ }\href@noop {} { (\bibinfo {year} {2008})},\ \Eprint
  {http://arxiv.org/abs/0812.1931} {arXiv:0812.1931 [astro-ph]} \BibitemShut
  {NoStop}%
%%CITATION = 0812.1931;%%
\bibitem [{\citenamefont {Arkani-Hamed}\ \emph {et~al.}(2009)\citenamefont
  {Arkani-Hamed}, \citenamefont {Finkbeiner}, \citenamefont {Slatyer},\ and\
  \citenamefont {Weiner}}]{ArkaniHamed:2008qn}%
  \BibitemOpen
  \bibfield  {author} {\bibinfo {author} {\bibfnamefont {N.}~\bibnamefont
  {Arkani-Hamed}}, \bibinfo {author} {\bibfnamefont {D.~P.}\ \bibnamefont
  {Finkbeiner}}, \bibinfo {author} {\bibfnamefont {T.~R.}\ \bibnamefont
  {Slatyer}}, \ and\ \bibinfo {author} {\bibfnamefont {N.}~\bibnamefont
  {Weiner}},\ }\Doi {10.1103/PhysRevD.79.015014} {\bibfield  {journal}
  {\bibinfo  {journal} {Phys. Rev.},\ }\textbf {\bibinfo {volume} {D79}},\
  \bibinfo {pages} {015014} (\bibinfo {year} {2009})},\ \Eprint
  {http://arxiv.org/abs/0810.0713} {arXiv:0810.0713 [hep-ph]} \BibitemShut
  {NoStop}%
%%CITATION = 0810.0713;%%
\bibitem [{\citenamefont {Baumgart}\ \emph {et~al.}(2009)\citenamefont
  {Baumgart}, \citenamefont {Cheung}, \citenamefont {Ruderman}, \citenamefont
  {Wang},\ and\ \citenamefont {Yavin}}]{Baumgart:2009tn}%
  \BibitemOpen
  \bibfield  {author} {\bibinfo {author} {\bibfnamefont {M.}~\bibnamefont
  {Baumgart}}, \bibinfo {author} {\bibfnamefont {C.}~\bibnamefont {Cheung}},
  \bibinfo {author} {\bibfnamefont {J.~T.}\ \bibnamefont {Ruderman}}, \bibinfo
  {author} {\bibfnamefont {L.-T.}\ \bibnamefont {Wang}}, \ and\ \bibinfo
  {author} {\bibfnamefont {I.}~\bibnamefont {Yavin}},\ }\Doi
  {10.1088/1126-6708/2009/04/014} {\bibfield  {journal} {\bibinfo  {journal}
  {JHEP},\ }\textbf {\bibinfo {volume} {04}},\ \bibinfo {pages} {014} (\bibinfo
  {year} {2009})},\ \Eprint {http://arxiv.org/abs/0901.0283} {arXiv:0901.0283
  [hep-ph]} \BibitemShut {NoStop}%
%%CITATION = 0901.0283;%%
\bibitem [{\citenamefont {Cui}\ \emph {et~al.}(2009)\citenamefont {Cui},
  \citenamefont {Morrissey}, \citenamefont {Poland},\ and\ \citenamefont
  {Randall}}]{Cui:2009xq}%
  \BibitemOpen
  \bibfield  {author} {\bibinfo {author} {\bibfnamefont {Y.}~\bibnamefont
  {Cui}}, \bibinfo {author} {\bibfnamefont {D.~E.}\ \bibnamefont {Morrissey}},
  \bibinfo {author} {\bibfnamefont {D.}~\bibnamefont {Poland}}, \ and\ \bibinfo
  {author} {\bibfnamefont {L.}~\bibnamefont {Randall}},\ }\href@noop {} {
  (\bibinfo {year} {2009})},\ \Eprint {http://arxiv.org/abs/0901.0557}
  {arXiv:0901.0557 [hep-ph]} \BibitemShut {NoStop}%
%%CITATION = 0901.0557;%%
\bibitem [{\citenamefont {Cheung}\ \emph {et~al.}(2009)\citenamefont {Cheung},
  \citenamefont {Ruderman}, \citenamefont {Wang},\ and\ \citenamefont
  {Yavin}}]{Cheung:2009qd}%
  \BibitemOpen
  \bibfield  {author} {\bibinfo {author} {\bibfnamefont {C.}~\bibnamefont
  {Cheung}}, \bibinfo {author} {\bibfnamefont {J.~T.}\ \bibnamefont
  {Ruderman}}, \bibinfo {author} {\bibfnamefont {L.-T.}\ \bibnamefont {Wang}},
  \ and\ \bibinfo {author} {\bibfnamefont {I.}~\bibnamefont {Yavin}},\ }\Doi
  {10.1103/PhysRevD.80.035008} {\bibfield  {journal} {\bibinfo  {journal}
  {Phys. Rev.},\ }\textbf {\bibinfo {volume} {D80}},\ \bibinfo {pages} {035008}
  (\bibinfo {year} {2009})},\ \Eprint {http://arxiv.org/abs/0902.3246}
  {arXiv:0902.3246 [hep-ph]} \BibitemShut {NoStop}%
%%CITATION = 0902.3246;%%
\bibitem [{\citenamefont {Katz}\ and\ \citenamefont
  {Sundrum}(2009)}]{Katz:2009qq}%
  \BibitemOpen
  \bibfield  {author} {\bibinfo {author} {\bibfnamefont {A.}~\bibnamefont
  {Katz}}\ and\ \bibinfo {author} {\bibfnamefont {R.}~\bibnamefont {Sundrum}},\
  }\Doi {10.1088/1126-6708/2009/06/003} {\bibfield  {journal} {\bibinfo
  {journal} {JHEP},\ }\textbf {\bibinfo {volume} {06}},\ \bibinfo {pages} {003}
  (\bibinfo {year} {2009})},\ \Eprint {http://arxiv.org/abs/0902.3271}
  {arXiv:0902.3271 [hep-ph]} \BibitemShut {NoStop}%
%%CITATION = 0902.3271;%%
\bibitem [{\citenamefont {Finkbeiner}\ \emph {et~al.}(2009)\citenamefont
  {Finkbeiner}, \citenamefont {Slatyer}, \citenamefont {Weiner},\ and\
  \citenamefont {Yavin}}]{Finkbeiner:2009mi}%
  \BibitemOpen
  \bibfield  {author} {\bibinfo {author} {\bibfnamefont {D.~P.}\ \bibnamefont
  {Finkbeiner}}, \bibinfo {author} {\bibfnamefont {T.~R.}\ \bibnamefont
  {Slatyer}}, \bibinfo {author} {\bibfnamefont {N.}~\bibnamefont {Weiner}}, \
  and\ \bibinfo {author} {\bibfnamefont {I.}~\bibnamefont {Yavin}},\ }\Doi
  {10.1088/1475-7516/2009/09/037} {\bibfield  {journal} {\bibinfo  {journal}
  {JCAP},\ }\textbf {\bibinfo {volume} {0909}},\ \bibinfo {pages} {037}
  (\bibinfo {year} {2009})},\ \Eprint {http://arxiv.org/abs/0903.1037}
  {arXiv:0903.1037 [hep-ph]} \BibitemShut {NoStop}%
%%CITATION = 0903.1037;%%
\bibitem [{\citenamefont {Alves}\ \emph {et~al.}(2009)\citenamefont {Alves},
  \citenamefont {Behbahani}, \citenamefont {Schuster},\ and\ \citenamefont
  {Wacker}}]{Alves:2009nf}%
  \BibitemOpen
  \bibfield  {author} {\bibinfo {author} {\bibfnamefont {D.~S.~M.}\
  \bibnamefont {Alves}}, \bibinfo {author} {\bibfnamefont {S.~R.}\ \bibnamefont
  {Behbahani}}, \bibinfo {author} {\bibfnamefont {P.}~\bibnamefont {Schuster}},
  \ and\ \bibinfo {author} {\bibfnamefont {J.~G.}\ \bibnamefont {Wacker}},\
  }\href@noop {} { (\bibinfo {year} {2009})},\ \Eprint
  {http://arxiv.org/abs/0903.3945} {arXiv:0903.3945 [hep-ph]} \BibitemShut
  {NoStop}%
%%CITATION = 0903.3945;%%
\bibitem [{\citenamefont {Morrissey}\ \emph {et~al.}(2009)\citenamefont
  {Morrissey}, \citenamefont {Poland},\ and\ \citenamefont
  {Zurek}}]{Morrissey:2009ur}%
  \BibitemOpen
  \bibfield  {author} {\bibinfo {author} {\bibfnamefont {D.~E.}\ \bibnamefont
  {Morrissey}}, \bibinfo {author} {\bibfnamefont {D.}~\bibnamefont {Poland}}, \
  and\ \bibinfo {author} {\bibfnamefont {K.~M.}\ \bibnamefont {Zurek}},\ }\Doi
  {10.1088/1126-6708/2009/07/050} {\bibfield  {journal} {\bibinfo  {journal}
  {JHEP},\ }\textbf {\bibinfo {volume} {07}},\ \bibinfo {pages} {050} (\bibinfo
  {year} {2009})},\ \Eprint {http://arxiv.org/abs/0904.2567} {arXiv:0904.2567
  [hep-ph]} \BibitemShut {NoStop}%
%%CITATION = 0904.2567;%%
\bibitem [{\citenamefont {Anber}\ \emph {et~al.}(2009)\citenamefont {Anber},
  \citenamefont {Aydemir}, \citenamefont {Donoghue},\ and\ \citenamefont
  {Pais}}]{Anber:2009tz}%
  \BibitemOpen
  \bibfield  {author} {\bibinfo {author} {\bibfnamefont {M.~M.}\ \bibnamefont
  {Anber}}, \bibinfo {author} {\bibfnamefont {U.}~\bibnamefont {Aydemir}},
  \bibinfo {author} {\bibfnamefont {J.~F.}\ \bibnamefont {Donoghue}}, \ and\
  \bibinfo {author} {\bibfnamefont {P.}~\bibnamefont {Pais}},\ }\Doi
  {10.1103/PhysRevD.80.015012} {\bibfield  {journal} {\bibinfo  {journal}
  {Phys. Rev.},\ }\textbf {\bibinfo {volume} {D80}},\ \bibinfo {pages} {015012}
  (\bibinfo {year} {2009})},\ \Eprint {http://arxiv.org/abs/0905.4260}
  {arXiv:0905.4260 [hep-ph]} \BibitemShut {NoStop}%
%%CITATION = 0905.4260;%%
\bibitem [{\citenamefont {Arina}\ \emph {et~al.}(2009)\citenamefont {Arina},
  \citenamefont {Ling},\ and\ \citenamefont {Tytgat}}]{Arina:2009um}%
  \BibitemOpen
  \bibfield  {author} {\bibinfo {author} {\bibfnamefont {C.}~\bibnamefont
  {Arina}}, \bibinfo {author} {\bibfnamefont {F.-S.}\ \bibnamefont {Ling}}, \
  and\ \bibinfo {author} {\bibfnamefont {M.~H.~G.}\ \bibnamefont {Tytgat}},\
  }\Doi {10.1088/1475-7516/2009/10/018} {\bibfield  {journal} {\bibinfo
  {journal} {JCAP},\ }\textbf {\bibinfo {volume} {0910}},\ \bibinfo {pages}
  {018} (\bibinfo {year} {2009})},\ \Eprint {http://arxiv.org/abs/0907.0430}
  {arXiv:0907.0430 [hep-ph]} \BibitemShut {NoStop}%
%%CITATION = 0907.0430;%%
\bibitem [{\citenamefont {Chen}\ \emph {et~al.}(2009)\citenamefont {Chen},
  \citenamefont {Cline},\ and\ \citenamefont {Frey}}]{Chen:2009ab}%
  \BibitemOpen
  \bibfield  {author} {\bibinfo {author} {\bibfnamefont {F.}~\bibnamefont
  {Chen}}, \bibinfo {author} {\bibfnamefont {J.~M.}\ \bibnamefont {Cline}}, \
  and\ \bibinfo {author} {\bibfnamefont {A.~R.}\ \bibnamefont {Frey}},\ }\Doi
  {10.1103/PhysRevD.80.083516} {\bibfield  {journal} {\bibinfo  {journal}
  {Phys. Rev.},\ }\textbf {\bibinfo {volume} {D80}},\ \bibinfo {pages} {083516}
  (\bibinfo {year} {2009})},\ \Eprint {http://arxiv.org/abs/0907.4746}
  {arXiv:0907.4746 [hep-ph]} \BibitemShut {NoStop}%
%%CITATION = 0907.4746;%%
\bibitem [{\citenamefont {Kaplan}\ \emph {et~al.}(2009)\citenamefont {Kaplan},
  \citenamefont {Krnjaic}, \citenamefont {Rehermann},\ and\ \citenamefont
  {Wells}}]{Kaplan:2009de}%
  \BibitemOpen
  \bibfield  {author} {\bibinfo {author} {\bibfnamefont {D.~E.}\ \bibnamefont
  {Kaplan}}, \bibinfo {author} {\bibfnamefont {G.~Z.}\ \bibnamefont {Krnjaic}},
  \bibinfo {author} {\bibfnamefont {K.~R.}\ \bibnamefont {Rehermann}}, \ and\
  \bibinfo {author} {\bibfnamefont {C.~M.}\ \bibnamefont {Wells}},\ }\href@noop
  {} { (\bibinfo {year} {2009})},\ \Eprint {http://arxiv.org/abs/0909.0753}
  {arXiv:0909.0753 [hep-ph]} \BibitemShut {NoStop}%
%%CITATION = 0909.0753;%%
\bibitem [{\citenamefont {Kumar}\ \emph {et~al.}(2009)\citenamefont {Kumar},
  \citenamefont {Tucker-Smith},\ and\ \citenamefont {Weiner}}]{Kumar:2009sf}%
  \BibitemOpen
  \bibfield  {author} {\bibinfo {author} {\bibfnamefont {A.}~\bibnamefont
  {Kumar}}, \bibinfo {author} {\bibfnamefont {D.}~\bibnamefont {Tucker-Smith}},
  \ and\ \bibinfo {author} {\bibfnamefont {N.}~\bibnamefont {Weiner}},\
  }\href@noop {} { (\bibinfo {year} {2009})},\ \Eprint
  {http://arxiv.org/abs/0910.2475} {arXiv:0910.2475 [hep-ph]} \BibitemShut
  {NoStop}%
%%CITATION = 0910.2475;%%
\bibitem [{\citenamefont {Chang}\ \emph
  {et~al.}(2010){\natexlab{b}}\citenamefont {Chang}, \citenamefont {Lang},\
  and\ \citenamefont {Weiner}}]{Chang:2010pr}%
  \BibitemOpen
  \bibfield  {author} {\bibinfo {author} {\bibfnamefont {S.}~\bibnamefont
  {Chang}}, \bibinfo {author} {\bibfnamefont {R.~F.}\ \bibnamefont {Lang}}, \
  and\ \bibinfo {author} {\bibfnamefont {N.}~\bibnamefont {Weiner}},\
  }\href@noop {} { (\bibinfo {year} {2010}{\natexlab{b}})},\ \Eprint
  {http://arxiv.org/abs/1007.2688} {arXiv:1007.2688 [hep-ph]} \BibitemShut
  {NoStop}%
%%CITATION = 1007.2688;%%
\bibitem [{\citenamefont {Kopp}\ \emph {et~al.}(2010)\citenamefont {Kopp},
  \citenamefont {Schwetz},\ and\ \citenamefont {Zupan}}]{Kopp:2009qt}%
  \BibitemOpen
  \bibfield  {author} {\bibinfo {author} {\bibfnamefont {J.}~\bibnamefont
  {Kopp}}, \bibinfo {author} {\bibfnamefont {T.}~\bibnamefont {Schwetz}}, \
  and\ \bibinfo {author} {\bibfnamefont {J.}~\bibnamefont {Zupan}},\ }\Doi
  {10.1088/1475-7516/2010/02/014} {\bibfield  {journal} {\bibinfo  {journal}
  {JCAP},\ }\textbf {\bibinfo {volume} {1002}},\ \bibinfo {pages} {014}
  (\bibinfo {year} {2010})},\ \Eprint {http://arxiv.org/abs/0912.4264}
  {arXiv:0912.4264 [hep-ph]} \BibitemShut {NoStop}%
%%CITATION = 0912.4264;%%
\bibitem [{\citenamefont {Lopes}(2010)}]{Lopes:2010zz}%
  \BibitemOpen
  \bibfield  {author} {\bibinfo {author} {\bibfnamefont {M.~I.}\ \bibnamefont
  {Lopes}} (\bibinfo {collaboration} {ZEPLIN-III}),\ }\Doi
  {10.1088/1742-6596/203/1/012025} {\bibfield  {journal} {\bibinfo  {journal}
  {J. Phys. Conf. Ser.},\ }\textbf {\bibinfo {volume} {203}},\ \bibinfo {pages}
  {012025} (\bibinfo {year} {2010})}\BibitemShut {NoStop}%
%%CITATION = 00462,203,012025;%%
\bibitem [{\citenamefont {Kuhlen}\ \emph {et~al.}(2010)\citenamefont {Kuhlen}
  \emph {et~al.}}]{Kuhlen:2009vh}%
  \BibitemOpen
  \bibfield  {author} {\bibinfo {author} {\bibfnamefont {M.}~\bibnamefont
  {Kuhlen}} \emph {et~al.},\ }\Doi {10.1088/1475-7516/2010/02/030} {\bibfield
  {journal} {\bibinfo  {journal} {JCAP},\ }\textbf {\bibinfo {volume} {1002}},\
  \bibinfo {pages} {030} (\bibinfo {year} {2010})},\ \Eprint
  {http://arxiv.org/abs/0912.2358} {arXiv:0912.2358 [astro-ph.GA]} \BibitemShut
  {NoStop}%
%%CITATION = 0912.2358;%%
\bibitem [{\citenamefont {Lang}\ and\ \citenamefont
  {Weiner}(2010)}]{Lang:2010cd}%
  \BibitemOpen
  \bibfield  {author} {\bibinfo {author} {\bibfnamefont {R.~F.}\ \bibnamefont
  {Lang}}\ and\ \bibinfo {author} {\bibfnamefont {N.}~\bibnamefont {Weiner}},\
  }\href@noop {} { (\bibinfo {year} {2010})},\ \Eprint
  {http://arxiv.org/abs/1003.3664} {arXiv:1003.3664 [hep-ph]} \BibitemShut
  {NoStop}%
%%CITATION = 1003.3664;%%
\bibitem [{\citenamefont {Alves}\ \emph {et~al.}(2010)\citenamefont {Alves},
  \citenamefont {Lisanti},\ and\ \citenamefont {Wacker}}]{Alves:2010pt}%
  \BibitemOpen
  \bibfield  {author} {\bibinfo {author} {\bibfnamefont {D.~S.~M.}\
  \bibnamefont {Alves}}, \bibinfo {author} {\bibfnamefont {M.}~\bibnamefont
  {Lisanti}}, \ and\ \bibinfo {author} {\bibfnamefont {J.~G.}\ \bibnamefont
  {Wacker}},\ }\href@noop {} { (\bibinfo {year} {2010})},\ \Eprint
  {http://arxiv.org/abs/1005.5421} {arXiv:1005.5421 [hep-ph]} \BibitemShut
  {NoStop}%
%%CITATION = 1005.5421;%%
\bibitem [{\citenamefont {Vogelsberger}\ \emph {et~al.}(2008)\citenamefont
  {Vogelsberger} \emph {et~al.}}]{Vogelsberger:2008qb}%
  \BibitemOpen
  \bibfield  {author} {\bibinfo {author} {\bibfnamefont {M.}~\bibnamefont
  {Vogelsberger}} \emph {et~al.},\ }\href@noop {} { (\bibinfo {year} {2008})},\
  \Eprint {http://arxiv.org/abs/0812.0362} {arXiv:0812.0362 [astro-ph]}
  \BibitemShut {NoStop}%
%%CITATION = 0812.0362;%%
\bibitem [{\citenamefont {Kim}(2008)}]{Kim:2008zzn}%
  \BibitemOpen
  \bibfield  {author} {\bibinfo {author} {\bibfnamefont {S.~K.}\ \bibnamefont
  {Kim}} (\bibinfo {collaboration} {KIMS}),\ }\Doi
  {10.1088/1742-6596/120/4/042021} {\bibfield  {journal} {\bibinfo  {journal}
  {J. Phys. Conf. Ser.},\ }\textbf {\bibinfo {volume} {120}},\ \bibinfo {pages}
  {042021} (\bibinfo {year} {2008})}\BibitemShut {NoStop}%
%%CITATION = 00462,120,042021;%%
\bibitem [{\citenamefont {Ressell}\ and\ \citenamefont
  {Dean}(1997)}]{Ressell:1997kx}%
  \BibitemOpen
  \bibfield  {author} {\bibinfo {author} {\bibfnamefont {M.~T.}\ \bibnamefont
  {Ressell}}\ and\ \bibinfo {author} {\bibfnamefont {D.~J.}\ \bibnamefont
  {Dean}},\ }\Doi {10.1103/PhysRevC.56.535} {\bibfield  {journal} {\bibinfo
  {journal} {Phys. Rev.},\ }\textbf {\bibinfo {volume} {C56}},\ \bibinfo
  {pages} {535} (\bibinfo {year} {1997})},\ \Eprint
  {http://arxiv.org/abs/hep-ph/9702290} {arXiv:hep-ph/9702290} \BibitemShut
  {NoStop}%
%%CITATION = HEP-PH/9702290;%%
\bibitem [{Note1()}]{Note1}%
  \BibitemOpen
  \bibinfo {note} {The flux of Standard Model monopoles is strongly
  bounded~\cite {Amsler:2008zzb}. A monopole under some other, broken $U(1)$
  has to be confined, which complicates the analysis, and we leave it for
  future work.}\BibitemShut {Stop}%
\bibitem [{\citenamefont {Bagnasco}\ \emph {et~al.}(1994)\citenamefont
  {Bagnasco}, \citenamefont {Dine},\ and\ \citenamefont
  {Thomas}}]{Bagnasco:1993st}%
  \BibitemOpen
  \bibfield  {author} {\bibinfo {author} {\bibfnamefont {J.}~\bibnamefont
  {Bagnasco}}, \bibinfo {author} {\bibfnamefont {M.}~\bibnamefont {Dine}}, \
  and\ \bibinfo {author} {\bibfnamefont {S.~D.}\ \bibnamefont {Thomas}},\ }\Doi
  {10.1016/0370-2693(94)90830-3} {\bibfield  {journal} {\bibinfo  {journal}
  {Phys. Lett.},\ }\textbf {\bibinfo {volume} {B320}},\ \bibinfo {pages} {99}
  (\bibinfo {year} {1994})},\ \Eprint {http://arxiv.org/abs/hep-ph/9310290}
  {arXiv:hep-ph/9310290} \BibitemShut {NoStop}%
%%CITATION = HEP-PH/9310290;%%
\bibitem [{\citenamefont {Pospelov}\ and\ \citenamefont {ter
  Veldhuis}(2000)}]{Pospelov:2000bq}%
  \BibitemOpen
  \bibfield  {author} {\bibinfo {author} {\bibfnamefont {M.}~\bibnamefont
  {Pospelov}}\ and\ \bibinfo {author} {\bibfnamefont {T.}~\bibnamefont {ter
  Veldhuis}},\ }\Doi {10.1016/S0370-2693(00)00358-0} {\bibfield  {journal}
  {\bibinfo  {journal} {Phys. Lett.},\ }\textbf {\bibinfo {volume} {B480}},\
  \bibinfo {pages} {181} (\bibinfo {year} {2000})},\ \Eprint
  {http://arxiv.org/abs/hep-ph/0003010} {arXiv:hep-ph/0003010} \BibitemShut
  {NoStop}%
%%CITATION = HEP-PH/0003010;%%
\bibitem [{\citenamefont {Sigurdson}\ \emph {et~al.}(2004)\citenamefont
  {Sigurdson}, \citenamefont {Doran}, \citenamefont {Kurylov}, \citenamefont
  {Caldwell},\ and\ \citenamefont {Kamionkowski}}]{Sigurdson:2004zp}%
  \BibitemOpen
  \bibfield  {author} {\bibinfo {author} {\bibfnamefont {K.}~\bibnamefont
  {Sigurdson}}, \bibinfo {author} {\bibfnamefont {M.}~\bibnamefont {Doran}},
  \bibinfo {author} {\bibfnamefont {A.}~\bibnamefont {Kurylov}}, \bibinfo
  {author} {\bibfnamefont {R.~R.}\ \bibnamefont {Caldwell}}, \ and\ \bibinfo
  {author} {\bibfnamefont {M.}~\bibnamefont {Kamionkowski}},\ }\Doi
  {10.1103/PhysRevD.70.083501} {\bibfield  {journal} {\bibinfo  {journal}
  {Phys. Rev.},\ }\textbf {\bibinfo {volume} {D70}},\ \bibinfo {pages} {083501}
  (\bibinfo {year} {2004})},\ \Eprint {http://arxiv.org/abs/astro-ph/0406355}
  {arXiv:astro-ph/0406355} \BibitemShut {NoStop}%
%%CITATION = ASTRO-PH/0406355;%%
\bibitem [{\citenamefont {Gardner}(2009)}]{Gardner:2008yn}%
  \BibitemOpen
  \bibfield  {author} {\bibinfo {author} {\bibfnamefont {S.}~\bibnamefont
  {Gardner}},\ }\Doi {10.1103/PhysRevD.79.055007} {\bibfield  {journal}
  {\bibinfo  {journal} {Phys. Rev.},\ }\textbf {\bibinfo {volume} {D79}},\
  \bibinfo {pages} {055007} (\bibinfo {year} {2009})},\ \Eprint
  {http://arxiv.org/abs/0811.0967} {arXiv:0811.0967 [hep-ph]} \BibitemShut
  {NoStop}%
%%CITATION = 0811.0967;%%
\bibitem [{\citenamefont {Cho}\ \emph {et~al.}(2010)\citenamefont {Cho},
  \citenamefont {Huh}, \citenamefont {Kim}, \citenamefont {Kim},\ and\
  \citenamefont {Kyae}}]{Cho:2010br}%
  \BibitemOpen
  \bibfield  {author} {\bibinfo {author} {\bibfnamefont {W.~S.}\ \bibnamefont
  {Cho}}, \bibinfo {author} {\bibfnamefont {J.-H.}\ \bibnamefont {Huh}},
  \bibinfo {author} {\bibfnamefont {I.-W.}\ \bibnamefont {Kim}}, \bibinfo
  {author} {\bibfnamefont {J.~E.}\ \bibnamefont {Kim}}, \ and\ \bibinfo
  {author} {\bibfnamefont {B.}~\bibnamefont {Kyae}},\ }\Doi
  {10.1016/j.physletb.2010.02.081} {\bibfield  {journal} {\bibinfo  {journal}
  {Phys. Lett.},\ }\textbf {\bibinfo {volume} {B687}},\ \bibinfo {pages} {6}
  (\bibinfo {year} {2010})},\ \Eprint {http://arxiv.org/abs/1001.0579}
  {arXiv:1001.0579 [hep-ph]} \BibitemShut {NoStop}%
%%CITATION = 1001.0579;%%
\bibitem [{\citenamefont {An}\ \emph {et~al.}(2010)\citenamefont {An},
  \citenamefont {Chen}, \citenamefont {Mohapatra}, \citenamefont {Nussinov},\
  and\ \citenamefont {Zhang}}]{An:2010kc}%
  \BibitemOpen
  \bibfield  {author} {\bibinfo {author} {\bibfnamefont {H.}~\bibnamefont
  {An}}, \bibinfo {author} {\bibfnamefont {S.-L.}\ \bibnamefont {Chen}},
  \bibinfo {author} {\bibfnamefont {R.~N.}\ \bibnamefont {Mohapatra}}, \bibinfo
  {author} {\bibfnamefont {S.}~\bibnamefont {Nussinov}}, \ and\ \bibinfo
  {author} {\bibfnamefont {Y.}~\bibnamefont {Zhang}},\ }\href@noop {} {
  (\bibinfo {year} {2010})},\ \Eprint {http://arxiv.org/abs/1004.3296}
  {arXiv:1004.3296 [hep-ph]} \BibitemShut {NoStop}%
%%CITATION = 1004.3296;%%
\bibitem [{\citenamefont {Dreiner}\ \emph {et~al.}(2008)\citenamefont
  {Dreiner}, \citenamefont {Haber},\ and\ \citenamefont
  {Martin}}]{Dreiner:2008tw}%
  \BibitemOpen
  \bibfield  {author} {\bibinfo {author} {\bibfnamefont {H.~K.}\ \bibnamefont
  {Dreiner}}, \bibinfo {author} {\bibfnamefont {H.~E.}\ \bibnamefont {Haber}},
  \ and\ \bibinfo {author} {\bibfnamefont {S.~P.}\ \bibnamefont {Martin}},\
  }\href@noop {} { (\bibinfo {year} {2008})},\ \Eprint
  {http://arxiv.org/abs/0812.1594} {arXiv:0812.1594 [hep-ph]} \BibitemShut
  {NoStop}%
%%CITATION = 0812.1594;%%
\bibitem [{\citenamefont {Masso}\ \emph {et~al.}(2009)\citenamefont {Masso},
  \citenamefont {Mohanty},\ and\ \citenamefont {Rao}}]{Masso:2009mu}%
  \BibitemOpen
  \bibfield  {author} {\bibinfo {author} {\bibfnamefont {E.}~\bibnamefont
  {Masso}}, \bibinfo {author} {\bibfnamefont {S.}~\bibnamefont {Mohanty}}, \
  and\ \bibinfo {author} {\bibfnamefont {S.}~\bibnamefont {Rao}},\ }\Doi
  {10.1103/PhysRevD.80.036009} {\bibfield  {journal} {\bibinfo  {journal}
  {Phys. Rev.},\ }\textbf {\bibinfo {volume} {D80}},\ \bibinfo {pages} {036009}
  (\bibinfo {year} {2009})},\ \Eprint {http://arxiv.org/abs/0906.1979}
  {arXiv:0906.1979 [hep-ph]} \BibitemShut {NoStop}%
%%CITATION = 0906.1979;%%
\bibitem [{\citenamefont {Barger}\ \emph {et~al.}(2010)\citenamefont {Barger},
  \citenamefont {Keung},\ and\ \citenamefont {Marfatia}}]{Barger:2010gv}%
  \BibitemOpen
  \bibfield  {author} {\bibinfo {author} {\bibfnamefont {V.}~\bibnamefont
  {Barger}}, \bibinfo {author} {\bibfnamefont {W.-Y.}\ \bibnamefont {Keung}}, \
  and\ \bibinfo {author} {\bibfnamefont {D.}~\bibnamefont {Marfatia}},\
  }\href@noop {} { (\bibinfo {year} {2010})},\ \Eprint
  {http://arxiv.org/abs/1007.4345} {arXiv:1007.4345 [hep-ph]} \BibitemShut
  {NoStop}%
%%CITATION = 1007.4345;%%
\bibitem [{\citenamefont {Akimov}\ \emph {et~al.}(2010)\citenamefont {Akimov}
  \emph {et~al.}}]{Akimov:2010vk}%
  \BibitemOpen
  \bibfield  {author} {\bibinfo {author} {\bibfnamefont {D.~Y.}\ \bibnamefont
  {Akimov}} \emph {et~al.},\ }\href@noop {} { (\bibinfo {year} {2010})},\
  \Eprint {http://arxiv.org/abs/1003.5626} {arXiv:1003.5626 [hep-ex]}
  \BibitemShut {NoStop}%
%%CITATION = 1003.5626;%%
\bibitem [{\citenamefont {Collaboration}(2009)}]{Collaboration:2009xb}%
  \BibitemOpen
  \bibfield  {author} {\bibinfo {author} {\bibfnamefont {X.}~\bibnamefont
  {Collaboration}},\ }\href@noop {} { (\bibinfo {year} {2009})},\ \Eprint
  {http://arxiv.org/abs/0910.3698} {arXiv:0910.3698 [astro-ph.CO]} \BibitemShut
  {NoStop}%
%%CITATION = 0910.3698;%%
\bibitem [{\citenamefont {Holdom}(1986)}]{Holdom:1985ag}%
  \BibitemOpen
  \bibfield  {author} {\bibinfo {author} {\bibfnamefont {B.}~\bibnamefont
  {Holdom}},\ }\Doi {10.1016/0370-2693(86)91377-8} {\bibfield  {journal}
  {\bibinfo  {journal} {Phys. Lett.},\ }\textbf {\bibinfo {volume} {B166}},\
  \bibinfo {pages} {196} (\bibinfo {year} {1986})}\BibitemShut {NoStop}%
%%CITATION = PHLTA,B166,196;%%
\bibitem [{\citenamefont {Lisanti}\ and\ \citenamefont
  {Wacker}(2009)}]{Lisanti:2009am}%
  \BibitemOpen
  \bibfield  {author} {\bibinfo {author} {\bibfnamefont {M.}~\bibnamefont
  {Lisanti}}\ and\ \bibinfo {author} {\bibfnamefont {J.~G.}\ \bibnamefont
  {Wacker}},\ }\href@noop {} { (\bibinfo {year} {2009})},\ \Eprint
  {http://arxiv.org/abs/0911.4483} {arXiv:0911.4483 [hep-ph]} \BibitemShut
  {NoStop}%
%%CITATION = 0911.4483;%%
\bibitem [{\citenamefont {Duda}\ \emph {et~al.}(2007)\citenamefont {Duda},
  \citenamefont {Kemper},\ and\ \citenamefont {Gondolo}}]{Duda:2006uk}%
  \BibitemOpen
  \bibfield  {author} {\bibinfo {author} {\bibfnamefont {G.}~\bibnamefont
  {Duda}}, \bibinfo {author} {\bibfnamefont {A.}~\bibnamefont {Kemper}}, \ and\
  \bibinfo {author} {\bibfnamefont {P.}~\bibnamefont {Gondolo}},\ }\href@noop
  {} {\bibfield  {journal} {\bibinfo  {journal} {JCAP},\ }\textbf {\bibinfo
  {volume} {0704}},\ \bibinfo {pages} {012} (\bibinfo {year} {2007})},\ \Eprint
  {http://arxiv.org/abs/hep-ph/0608035} {arXiv:hep-ph/0608035} \BibitemShut
  {NoStop}%
%%CITATION = HEP-PH/0608035;%%
\bibitem [{\citenamefont {Ellis}\ and\ \citenamefont
  {Flores}(1991)}]{Ellis:1991ef}%
  \BibitemOpen
  \bibfield  {author} {\bibinfo {author} {\bibfnamefont {J.~R.}\ \bibnamefont
  {Ellis}}\ and\ \bibinfo {author} {\bibfnamefont {R.~A.}\ \bibnamefont
  {Flores}},\ }\Doi {10.1016/0370-2693(91)90597-J} {\bibfield  {journal}
  {\bibinfo  {journal} {Phys. Lett.},\ }\textbf {\bibinfo {volume} {B263}},\
  \bibinfo {pages} {259} (\bibinfo {year} {1991})}\BibitemShut {NoStop}%
%%CITATION = PHLTA,B263,259;%%
\bibitem [{\citenamefont {Toivanen}\ \emph {et~al.}(2009)\citenamefont
  {Toivanen}, \citenamefont {Kortelainen}, \citenamefont {Suhonen},\ and\
  \citenamefont {Toivanen}}]{Toivanen:2009zza}%
  \BibitemOpen
  \bibfield  {author} {\bibinfo {author} {\bibfnamefont {P.}~\bibnamefont
  {Toivanen}}, \bibinfo {author} {\bibfnamefont {M.}~\bibnamefont
  {Kortelainen}}, \bibinfo {author} {\bibfnamefont {J.}~\bibnamefont
  {Suhonen}}, \ and\ \bibinfo {author} {\bibfnamefont {J.}~\bibnamefont
  {Toivanen}},\ }\Doi {10.1103/PhysRevC.79.044302} {\bibfield  {journal}
  {\bibinfo  {journal} {Phys. Rev.},\ }\textbf {\bibinfo {volume} {C79}},\
  \bibinfo {pages} {044302} (\bibinfo {year} {2009})}\BibitemShut {NoStop}%
%%CITATION = PHRVA,C79,044302;%%
\end{thebibliography}%
\end{document}